\documentclass[conference]{IEEEtran}
\IEEEoverridecommandlockouts
\usepackage{cite}

\usepackage{amsmath,amssymb,amsfonts,amsthm}
\usepackage{algorithmic}
\usepackage{multirow,makecell}
\usepackage{graphicx}
\usepackage{textcomp}
\usepackage{xcolor}
\usepackage{hyperref}
\usepackage[hyphenbreaks]{breakurl}
\usepackage{stmaryrd}
\usepackage{graphpap}
\usepackage{graphicx} %use graph format
\usepackage{epstopdf}
\usepackage{indentfirst}
\usepackage{color}
\usepackage{tikz}
\usetikzlibrary{shapes,arrows}

\usepackage{prftree} 

\def\BibTeX{{\rm B\kern-.05em{\sc i\kern-.025em b}\kern-.08em
    T\kern-.1667em\lower.7ex\hbox{E}\kern-.125emX}}

\newcommand{\creturn}[1]{\textbf{return}\ #1}
\newcommand{\ereturn}{\textbf{retv}}
\newcommand{\vreturn}{\textbf{ret}}
\newcommand{\hP}{h^{\prime}}
\newcommand{\PHI}{\mathbf{\Phi}}

\newcommand{\pskip}{\textmd{skip}}

\newcommand{\evo}[3]{\langle \dot{#1}=#2\& #3\rangle}

\newcommand{\pwait}{\textrm{wait}}
\newcommand{\seman}[1]{[\![#1]\!]}

\newcommand{\mainthread}{\textbf{t}_m}

\newcommand{\exempt}[4]{#1 \unrhd \talloblong_{#2} (#3 \rightarrow #4)}

\newcommand{\oomit}[1]{}
\newcommand{\IFE}[3]{\textbf{if}\ #1\ #2\ \textbf{else}\ #3}

\newcommand{\for}[4]{\textbf{for}(#1, #2, #3)\ #4}
\newcommand{\compat}{\operatorname{compat}}
\newcommand{\rdy}{\mathit{rdy}}
 
\newcommand{\cwhile}[2]{\textbf{while}\ #1 \ #2}
\newcommand{\clock}[2]{\textbf{lock}\ #1\ #2}
\newcommand{\cunlock}[2]{\textbf{unlock}\ #1\ #2}
\newcommand{\cwait}[3]{\textbf{cwait}\ #1\ #2\ #3}
\newcommand{\lock}[1]{\textbf{lock}\ #1}
\newcommand{\unlock}[1]{\textbf{unlock}\ #1}
\newcommand{\wait}[2]{\textbf{cwait}\ #1\ #2}

\newcommand{\signal}[1]{\textbf{signal}\ #1}
\newcommand{\create}[3]{\textbf{create}\ #1\ #2\ #3}
\newcommand{\join}[1]{\textbf{join}\ #1}

\newcommand{\toc}{\textit{HtoC}}
\newcommand{\DHCSP}{\textit{HtoD}}

\newcommand{\sm}[1]{$#1$}

\newcommand{\qq}{q}

\newcommand{\xx}{\mathbf{x}}
\newcommand{\yy}{\mathbf{y}}
\newcommand{\ff}{\mathbf{f}}
\newcommand{\RR}{\mathbb{R}}
\newcommand{\euler}{\xi}

	\tikzstyle{startstop} = [rectangle,rounded corners, minimum width=3cm,minimum height=1cm,text centered,text width =3cm, draw=black]
	\tikzstyle{io} = [trapezium, trapezium left angle = 70,trapezium right angle=110,minimum width=3cm,minimum height=1cm,text centered,text width =3cm,draw=black]
	\tikzstyle{process} = [rectangle,minimum width=3cm,minimum height=1cm,text centered,text width =3cm,draw=black]
	\tikzstyle{decision} = [diamond,aspect = 3.5,text centered,draw=black]
	\tikzstyle{arrow} = [thick,->,>=stealth]
	\tikzstyle{straightline} = [line width = 1pt,-]
	\tikzstyle{point}=[coordinate]
	\tikzset{global scale/.style={
			scale=#1,
			every node/.append style={scale=#1}
		}
	}

\newcommand{\newcomment}[1]{#1}
\newtheorem{definition}{Definition}
\newtheorem{proposition}{Proposition}
\newtheorem{theorem}{Theorem}

\usepackage{listings}
\usepackage{xcolor}

\definecolor{codegreen}{rgb}{0,0.6,0}
\definecolor{codegray}{rgb}{0.5,0.5,0.5}
\definecolor{codepurple}{rgb}{0.58,0,0.82}
\definecolor{backcolour}{rgb}{0.95,0.95,0.92}

\lstdefinestyle{mystyle}{
    backgroundcolor=\color{backcolour},   
    commentstyle=\color{codegreen},
    keywordstyle=\color{magenta},
    numberstyle=\tiny\color{codegray},
    stringstyle=\color{codepurple},
    basicstyle=\ttfamily\scriptsize,
    breakatwhitespace=false,         
    breaklines=true,                 
    captionpos=b,                    
    keepspaces=true,                 
    numbers=left,                    
    numbersep=5pt,                  
    showspaces=false,                
    showstringspaces=false,
    showtabs=false,                  
    tabsize=2,
    morekeywords={lock, unlock}
}

\lstdefinelanguage{custom}
{
language=C,
morekeywords={lock, unlock, cwait, signal, join, creat, Channel}
}

\begin{document}

\title{Formally Verified C Code Generation from Hybrid Communicating Sequential Processes (Full Version)\\
%\thanks{Supported by ...}
}

\author{\IEEEauthorblockN{1\textsuperscript{st} Shuling Wang}
 \IEEEauthorblockA{\textit{State Key Lab. of Computer Science} \\
 \textit{Institute of Software, Chinese Academy of Sciences}\\
 Beijing, China \\
 wangsl@ios.ac.cn}
 \and
 \IEEEauthorblockN{2\textsuperscript{nd} Zekun Ji}
 \IEEEauthorblockA{\textit{State Key Lab. of Computer Science} \\
 \textit{Institute of Software, Chinese Academy of Sciences}\\
 Beijing, China \\
 jizk@ios.ac.cn}
 \and
 \IEEEauthorblockN{3\textsuperscript{rd} Bohua Zhan}
 \IEEEauthorblockA{\textit{State Key Lab. of Computer Science} \\
 \textit{Institute of Software, Chinese Academy of Sciences}\\
 Beijing, China \\
 bzhan@ios.ac.cn}
 \and
 \IEEEauthorblockN{4\textsuperscript{th} Xiong Xu}
 \IEEEauthorblockA{\textit{State Key Lab. of Computer Science} \\
 \textit{Institute of Software, Chinese Academy of Sciences}\\
 Beijing, China \\
 xux@ios.ac.cn}
 \and
 \IEEEauthorblockN{5\textsuperscript{th} Qiang Gao}
 \IEEEauthorblockA{\textit{State Key Lab. of Computer Science} \\
 \textit{Institute of Software, Chinese Academy of Sciences}\\
 Beijing, China \\
 gaoq@ios.ac.cn}
 \and
 \IEEEauthorblockN{6\textsuperscript{th} Naijun Zhan}
 \IEEEauthorblockA{\textit{State Key Lab. of Computer Science} \\
 \textit{Institute of Software, Chinese Academy of Sciences}\\
 Beijing, China \\
 znj@ios.ac.cn}
 }

\maketitle

\begin{abstract}
Hybrid Communicating Sequential Processes (HCSP) is a formal model for hybrid systems, including primitives for evolution along an ordinary differential equation (ODE), communication, and parallel composition. Code generation is needed to convert HCSP models into code that can be executed in practice, and the correctness of this conversion is essential to ensure that the generated code accurately reflects the formal model. In this paper, we propose a code generation algorithm from HCSP to C with POSIX library for concurrency. The main difficulties include how to bridge the gap between the synchronized communication model in HCSP and the use of mutexes for synchronization in C, and how to discretize evolution along ODEs and support interrupt of ODE evolution by communication. To prove the correctness of code generation, we define a formal semantics for POSIX C, and build transition system models for both HCSP and C programs. We then define an approximate bisimulation relation between traces of transition systems, and show that under certain robustness conditions for HCSP, the generated C program is approximately bisimilar to the original model. Finally, we evaluate the code generation algorithm on a detailed model for automatic cruise control, showing its utility on real-world examples.
\end{abstract}


\section{Introduction}

Cyber-Physical Systems (CPSs) can be complex, networked, systems of systems, and are often entrusted with safety-critical tasks. The efficient and verified development of safe and reliable CPSs is a priority mandated by many standards, yet a notoriously difficult and challenging field.
To address design complexity under the necessity of verified CPS development, Model-Driven Development (MDD) has become a predominant approach in CPS development. \newcomment{MDD usually comprises different abstraction levels from top to bottom, e.g. graphical models, formal models, and code at the implementation level. }There are two orthogonal principles followed by MDD:
\begin{itemize}
\item \textbf{Horizontal} Composition/Decomposition: $P\|Q$.
\item \textbf{Vertical} Abstraction/Refinement: $P\sqsubseteq Q$.
\end{itemize}

\newcomment{The horizontal dimension requires a modelling mechanism that is compositional such that the complexity of modelling and verifying the system can be reduced by separately verifying its subcomponents. On the vertical dimension, specific modelling and analysis are performed at each abstraction level, and each level is refined by the more concrete level so that the behaviors are preserved.} Therefore, if formal models are proved correct, the code generated from them is also guaranteed to be correct without further proof.

\newcomment{
 Especially, the MDD of hybrid systems that integrate traditional discrete models with dynamic models faces a central problem: how
to transform an abstract hybrid control model to an algorithmic model at code level rigorously
and automatically. The controller code determines how to sample data
from the continuous plant and the entanglement between sampling data and computing control
commands is intricate. An efficient approach is to discretize the continuous plant, and then, the
discretized continuous plant together with the controller code constitute an embedded real-time implementation for  the closed-loop hybrid system. How and according to which criterion to discretize
 the continuous behaviour and then generating correct C code, is addressed in this paper. }

There are many industrial MDD tools targeting CPS design and development, such as Simulink/Stateflow~\cite{slusing,sfusing}, SCADE~\cite{dormoy2008scade} and so on.  Simulink/Stateflow and SCADE both support automatic code generation from control models, targeting at real-time applications. However, Simulink/Stateflow can only guarantee correctness of generated code by (incomplete) simulations. \newcomment{SCADE was founded on the synchronous dataflow language Lustre~\cite{lustre}, and its formally verified compiler V\'elus~\cite{PLDI17} guarantees that the generated  code faithfully implements the  
semantics of Lustre. However, it lacks support for continuous plants modelling.}  In the academic community, there are also a number of studies on formal modelling and verification of CPSs, for instance, hybrid and timed automata~\cite{DBLP:conf/lics/Henzinger96,DBLP:conf/cav/FengKLXZ18,DBLP:journals/tcad/WangZA18}, hybrid programs and dynamic differential logic~\cite{DBLP:books/sp/Platzer18}, Event-B and its hybridation~\cite{DBLP:journals/tecs/DupontASP21}, and hybrid process algebra~\cite{DBLP:journals/jlp/CuijpersR05,DBLP:journals/tcs/BergstraM05}. Most of these approaches address verification only, while Event-B supports code generation with formal guarantee but only for discrete case. 

Hybrid CSP (HCSP)~\cite{HCSP1994,Zhan2017book}, a hybrid extension to the classic CSP (Communicating Sequential Processes)~\cite{hoare1978communicating}, is a compositional formalism in describing hybrid systems. It uses ordinary differential equations (ODEs) to model continuous evolutions and introduces interrupts to model interactions between continuous and discrete dynamics in a very flexible manner. The verification of HCSP is conducted by tools based on Hybrid Hoare Logic (HHL)~\cite{Zhan23}, including verification within the interactive theorem prover Isabelle/HOL~\cite{ZouZWFQ13,Zhan2017book} as well as a more automatic tool $\mathsf{HHLPy}$~\cite{DBLP:conf/fm/ShengBZ23}.
Graphical models for CPSs can be translated into HCSP~\cite{ZouZWFQ13,DBLP:journals/tcs/XuWZJTZ22}, and the verified HCSP can then be transformed to executable SystemC code\cite{DBLP:journals/tosem/YanJWWZ20}. To guarantee the correctness of transformation from HCSP to SystemC, the notion of approximate bisimulation proposed by~\cite{DBLP:journals/tac/GirardP07,DBLP:journals/scl/JuliusDBP09} was used to measure the equivalence between hybrid and discrete systems, to allow a distance between the observations of two systems within a tolerable bound rather than exactly identical, which was proved to hold for HCSP and SystemC under some \newcomment{robustly-safety} conditions in~\cite{DBLP:journals/tosem/YanJWWZ20}.  

\newcomment{This paper builds upon the work of Yan et al. in~\cite{DBLP:journals/tosem/YanJWWZ20}. It aims at generating code in C from HCSP models using the concurrency primitives of the POSIX $\mathsf{pthreads}$ library, with correctness guarantees. 
It adopts the notions of approximate bisimulation, robustly-safety, and some of the discretisation rules of HCSP proposed in~\cite{DBLP:journals/tosem/YanJWWZ20}. However, it is distinguished from previous work in the following aspects:
\begin{enumerate}
 \item Although both works transform HCSP to its discretized version and then to code, in~\cite{DBLP:journals/tosem/YanJWWZ20}, the discretization of the HCSP constructs related to communications, including inputs, outputs and continuous interrupt, is defined with the help of shared channel variables indicating their readiness to perform communication actions (see Table 2 and 3 of~\cite{DBLP:journals/tosem/YanJWWZ20});  and this work does not need to introduce extra shared variables at this step, thus keeps communications the same as original HCSP. 
 \item SystemC supports communication mechanisms similar to HCSP on its own right, so the communications in HCSP can be translated naturally to SystemC. Compared to this, communications are implemented in C using the POSIX $\mathsf{pthreads}$ library, in terms of mutexes and condition variables to achieve time and value synchronization. Hence, the method of translation to C is more involved. At semantic level,  concurrency in HCSP follows a hand-shaking model while  in C it follows an interleaving model controlled by mutexes and condition variables. Thus, the two concurrency mechanisms in HCSP and C are completely different, and how to prove the equivalence between them is one of the main challenges addressed in our work.
\end{enumerate} 
We prove the approximate bisimulation between an HCSP process and the generated C code in two steps:  approximate bisimulation between the HCSP process and its discretised version, and bisimulation between the discrete HCSP process and the generated C code. Compared to the proofs of~\cite{DBLP:journals/tosem/YanJWWZ20}, the first part proves the case for continuous evolution and the case for continuous interrupt based on its new discretization, and the second part is completely new: we introduce a new bisimulation relation, for which each transition step in HCSP may correspond to multiple atomic blocks of execution on the C side, but only one of them is considered as the essential step to perform the transition.}

\oomit{Added\color{red} Introduction still to add:
\begin{itemize}
\item Concept of bisimulation and approximate bisimulation.
\item What are the main challenges for translation and correctness proof? (difference between the synchronization model of concurrency in HCSP and interleaving model in C).
\item Outline of correctness proof: define simulation relation, each transition step in HCSP may correspond to multiple atomic blocks of execution on the C side, but only one of them is considered to perform the transition.
\end{itemize}
}
In summary, the main contributions of this paper comprise\footnote{The source code and examples of the code generator are available at \url{https://github.com/jimoc2048bits/CCodeGenerationFromHCSP}.}:
\begin{enumerate}
\item We present a formal semantics for a subset of C language with POSIX threads.
\item \newcomment{We implement the synchronized communications of HCSP with the use of mutexes in the $\mathsf{pthreads}$ library, based on which we realise the transformation from any HCSP process to C code. }
\item Based on the notions of approximate bisimulation, we prove the correctness of the transformation from HCSP to C. The proof uses a new bisimulation relation for verifying equivalence between two concurrency mechanisms with synchronized and interleaving settings. 
\item We apply our approach on a realistic Automatic Cruise Control System, including its HCSP model, the C code generated from the   model, and the comparison with original HCSP and other C implementation by simulation. 
\end{enumerate}

After reviewing related works, the paper is organized as follows: Sect.~\ref{Preliminaries} introduces some preliminary knowledge of this work. Sect.~\ref{Target} introduces the syntax and semantics of C with POSIX threads. The translation from HCSP to C is specified in Sect.~\ref{HCSP2C}, and  the correctness of the translation is justified in Sect.~\ref{sec:correct}. Sect.~\ref{CaseStudy} illustrates our approach by a realistically-scaled case study. Sect.~\ref{Conclusions} concludes.

\subsection{Related Work}

\newcomment{Model-based automatic code generation has been extensively studied in both academic and industrial communities~\cite{DBLP:journals/tc/AnandFHKL10}, but code generation that supports hybrid systems and provides formal correctness guarantees of generated code at the same time is not well addressed. Some examples include the aforementioned Simulink~\cite{slusing}, SCADE~\cite{dormoy2008scade} founded on synchronous Lustre, and the formal modelling languages~\cite{DBLP:conf/lics/Henzinger96,DBLP:conf/cav/FengKLXZ18,DBLP:journals/tcad/WangZA18}.  OSATE/AADL~\cite{OSATE} provides architecture modeling and analysis of real-time systems  and furthermore supports the
automated code generation from AADL models including runtime behavior and scheduling to C code. However, it
 validates the code generation by simulation, and moreover  does not support
continuous time modeling. 
The compiler Z\'elus~\cite{HSCC2013} extends  Lustre~\cite{lustre}  with ODEs and implements code generation from the extended hybrid language, which has also been implemented in SCADE 6.  It supports analysis of hybrid models by type systems and semantics, and handles the detection of zero-crossing events~\cite{HSCC2013,JCSS2012}. But it does not explicitly support constructs related to communication and concurrency. 
VeriPhy~\cite{DBLP:conf/pldi/BohrerTMMP18} automatically transforms verified formal  models of  CPSs modelled in differential dynamic logic (dL)~\cite{DBLP:books/sp/Platzer18,DBLP:conf/cade/FultonMQVP15} to controller implementations that preserve safety properties of original models rather than their semantics. Thus compared to our work, it does not consider the (approximate) equivalence between the source models with ODEs and the discrete implementation, and the zero-crossing problems caused by discretization. }

%Discretization of continuous models.

%From hybrid models to executable code.

%Code generation tools.

\section{Preliminaries}\label{Preliminaries}
This section introduces the notion of transition systems, approximate bisimulation, discretization of ODEs, and HCSP.  

\subsection{Transition Systems}
\label{sec:tsandab}

\begin{definition}[Transition system]
A transition system  is a tuple $T=\langle Q,L,\to,Q^{0}, Y, H\rangle $, where $Q$ is a  set of states, $L \subseteq \mathcal{ACT} \cup \{\tau\}$ is a  set of labels, $\to\,\subseteq Q \times L \times Q$ is a set of transitions,  $Q^{0}\subseteq Q$ is a set of initial states, $Y$ is a set of observations, and $H: Q \rightarrow Y$ is an observation function. $\mathcal{ACT}$ is a set of events and $\tau$ is an internal event  ($\tau \notin \mathcal{ACT}$).  
\label{definition:TS}
\end{definition}
Given a transition system, for any $a\in \mathcal{ACT}$, we define the $\tau$-closed transition $q\stackrel{a}{\Rightarrow} q'$ to represent that $q$ can reach $q'$ via action $a$ and a sequence of $\tau$ actions, i.e.\newcomment{$q (\xrightarrow{\tau})^* q_i\xrightarrow{a} q_{i+1}(\xrightarrow{\tau})^* q'$}. %We can define trajectories of transitions for both $\rightarrow$ and $\Rightarrow$. For the first case, a  \emph{trajectory} of a transition system $T$ is a  possibly infinite sequence of transitions $q^0 \xrightarrow{l^{0}} q^1 \xrightarrow{l^{1}}\cdots \xrightarrow{l^{i-1}} q^{i}\xrightarrow{l^{i}} \cdots$, denoted by $\{q^i\xrightarrow{l^i}q^{i+1}\}_{i\in \mathbb{N}}$,  s.t.  $q^{0}\in Q^0$ and for any $i$,  $q^{i} \xrightarrow{l^{i}} q^{i+1}$. The trajectories for $\tau$-closed transitions can be obtained by replacing $\rightarrow$ with $\Rightarrow$. Given two events $l_1, l_2$, we define the distance between them, denoted by $\labeld(l_1, l_2)$. 
We will define the semantics of HCSP and C using transition systems. \newcomment{In many cases we set $\mathcal{ACT} = \emptyset$ (such as when modelling the semantics for C), then we will define $\to\ \subseteq Q \times Q$ instead, and $\Rightarrow$ is equivalent to $\rightarrow^*$.  }

\subsection{Approximate Bisimulation}

The notion of approximate bisimulation was first proposed in~\cite{DBLP:journals/tac/GirardP07} to measure the equivalence between hybrid systems, to allow a limited distance between the observations of two systems. Later in~\cite{DBLP:journals/scl/JuliusDBP09},  it was extended to allow  precision not only between the observations, but also between the synchronisation labels  of two systems.  In~\cite{DBLP:journals/tosem/YanJWWZ20}, the authors instantiate the precision parameters to be the time and value tolerances between two systems. Below we present the notion of approximate bisimulation in~\cite{DBLP:journals/tosem/YanJWWZ20}. 
%A modification of this definition will be given in Sect.~\ref{sec:correct} in order to meet our requirement. 
%The approximate bisimulation is defined between two transition systems. 
Let $TS_{i}=\langle Q_i,L_i,\rightarrow_i,Q^{0}_i, Y_i, H_i\rangle $, $(i=1,2)$ be two transition systems, $h$ and $\varepsilon$ the time and value precisions  resp.
%$\textit{Vars}(\cdot)$  the set of variables occurring in current process or code. 

\begin{definition}[Approximate Bisimulation]
 $\mathcal{B}_{h, \varepsilon}\subseteq Q_{1}\times Q_{2}$ is called a $(h, \varepsilon)$-approximate bisimulation relation between $TS_{1}$ and $TS_{2}$, if it is symmetric, and for all $(q_{1},q_{2})\in \mathcal{B}_{h,\varepsilon}$, 
 \begin{itemize}
     \item The distance between the observations is within the given value precision, i.e.  $|H_1(q_{1}),H_2(q_{2})| \le \varepsilon$, \newcomment{where $|H_1(q_{1}),H_2(q_{2})|$ returns the maximum of Euclidean distances of observation variables in $q_1$ and $q_2$}.

     \item $ \forall \qq_{1} \stackrel{l_1}{\rightarrow}_1 \qq_{1}'$, $\exists \qq_{2} \stackrel{ l_2}{\Rightarrow}_2 \qq_{2}'$  s.t.  $(\qq'_{1},\qq'_{2})\in \mathcal{B}_{h, \varepsilon}$, and $|l_{1},l_{2}| < h$, for $l_1\in L_1, l_2 \in L_2$.  \newcomment{Here $|l_{1},l_{2}|$ is 0 if $l_1=l_2$, $|l_1-l_2|$ if $l_1, l_2 \in R$, and $\infty$ otherwise.  } 
 \end{itemize}
\label{definition:appbisimulation}
\end{definition}

The $(h, \varepsilon)$-approximate bisimulation requires the distance between the observations under the pair $(q_1, q_2)$ must be within $\varepsilon$; 
furthermore, if one of them is able to reach a state via an event, the other one can
also reach a state via an event such that the distance between the events is within $h$ and
the pair of resulting states also satisfy the approximate bisimulation.  
%The bisimulation of two transition systems is given as follows:
\begin{definition}
  $\mathcal{TS}_1$ and $\mathcal{TS}_2$ are approximate bisimilar \newcomment{with respect to $h$ and $ \varepsilon$}, denoted $\mathcal{TS}_1 \cong_{h, \varepsilon} \mathcal{TS}_2$, if there exists a  bisimulation relation $\mathcal{B}_{h, \varepsilon}$ satisfying that for all initial configurations $q_1 \in {Q_1}^0$ there exists $q_2 \in {Q_2}^0$ such that $(q_1, q_2) \in \mathcal{B}_{h, \varepsilon}$ and vice versa. 
\end{definition}

\subsection{Discretization of ODE}
Here we present the discretization of ODEs and its correctness, which have been studied in ~\cite{DBLP:journals/tosem/YanJWWZ20}. We briefly revisit the related results in~\cite{DBLP:journals/tosem/YanJWWZ20} here. 

We apply the 4-stage Runge-Kutta method to discretize the continuous dynamics, which is more effective with  \emph{global discretization error} $O(h^4)$. The ODE $\dot{\xx} = \ff(\xx) $ on $[t_0, t_0+T_o]$ is discretized as
\begin{equation*}
(\pwait\ h; \xx:= \xx+h\PHI(\xx, h))^N;\pwait\ \hP; \xx:= \xx+\hP\PHI(\xx, h')
\end{equation*}
where $N= \lfloor \frac{T_o}{h}\rfloor$, $\hP=T_o-Nh$, and $\PHI(\xx, s)=\frac{1}{6}(k_1+2k_2+2k_3+k_4)$ with $k_1=\ff(\xx)$,
$k_2=\ff(\xx+\frac{1}{2}sk_1)$,
$k_3=\ff(\xx+\frac{1}{2}sk_2)$ and
$k_4=\ff(\xx+sk_3)$.
With the initial state $\xx_0$ at $h_0=t_0$, the obtained sequence of approximate solutions $\{\xx_i\}$ at time stamps $\{h_i\}$ is (below $1\le j\le N$):
\begin{equation*}
\label{Def:handx}
\left\{
\begin{array}{lll}
   \xx_0, & h_0=t_0, & \\
   \xx_j = \xx_{j-1} + h \PHI(\xx_{j-1}, h), & h_j = t_0 + j*h \\
   \xx_{N+1} = \xx_N + \hP \PHI(\xx_N, h'), & h_{N+1} = t_0+T_o
\end{array}
\right.
\end{equation*}
Intuitively,
$T_o$ is divided into $N$ intervals of length $h$ and a possible residual  interval of length $\hP$.
$\PHI$ is computed based on the values of the vector field at the four points and used for approximating the value of $\xx$.
\newcomment{Below we present the global error of the discretization (see Theorem 7.2.2.3 in~\cite{globalerror})}.
\begin{proposition}[Global Error]
   \newcomment{Assume the ODE $\dot{\xx} = \ff(\xx)$ satisfies the local Lipschitz condition,  that is,  for any compact set $S$ of $\RR^n$, $\|\ff(\yy_1) - \ff(\yy_2)\| \leq L\|\yy_1 - \yy_2\|$ for all $\yy_1, \yy_2 \in S$.} Let $X(t, \widetilde{\xx}_0)$ be the exact  solution of the ODE with initial value $\widetilde{\xx}_0$  on $[0, T_0]$.  Suppose $\xx_0 \in \RR^n$ is a state with $\|\xx_0 - \widetilde{\xx}_0\| \le \euler_1$. Then there exists a discretized step $h_e > 0$ s.t. for all $0<h\leq h_e$ and all $ i \leq \lceil \frac{T_o}{h} \rceil$,
  %and $i\le N+1$ with $N=\lfloor \frac{T_o}{h}\rfloor$,
  the global discretization error between $X(h_i, \widetilde{\xx}_0)$ and $\xx_i$ satisfies:
   \begin{equation*}   \begin{array}{ll}
   \|X(h_i, \widetilde{\xx}_0) - \xx_i\| \leq  M(h),
  \mbox{  where } \\
     M(h) = \frac{e^{L\hP} -1}{L} C_2 (\hP)^4+[1+L\hP+\frac{(L\hP)^2}{2}+\frac{(L\hP)^3}{4}
     \\
     \qquad \qquad +\frac{(L\hP)^4}{24}]M_N(h) \\ 
     M_N(h)= e^{NLh}\euler_1 + \frac{e^{NLh} -1}{L}C_1h^4.
    % \hP=& T_o-Nh
    \end{array}
   \end{equation*}
Among them  $N,\hP, h_i$ and $\xx_i$ are as defined previously, and $C_1$, $C_2$ are positive constants depending on the local discretization error of the 4-stage Runge-Kutta method. \newcomment{Here given a vector $\xx \in \mathbb{R}^n$, $\| \xx\|$ denotes the infinity norm of $\xx$, i.e., $\| \xx \|=\max\{|x_1|,|x_2|,...,|x_n|\}$.}
%   \vspace*{-5mm}
\label{theorem:globalerror}
\end{proposition}

From the definition of $M(h)$, the global error is monotonically increasing with respect to step size $h$, the Lipschitz constant $L$, and the two local discretization error constants $C_1$ and $C_2$. 
\oomit{We will use this fact to compute the step $h$ for discretizing a HCSP process that may contain multiple different ODEs. }
\newcomment{In~\cite{DBLP:journals/tosem/YanJWWZ20},  the following theorem is proved for correctness of the discretization of an ODE.}
\oomit{The transition systems for judging the approximate bisimulation between the ODE and its discretization are similar to the corresponding constructs in HCSP transition semantics and neglected here. }

\begin{theorem}[Correctness of Discretization of ODEs] 
Suppose $L$ is the Lipschitz constant of $\dot{\xx} = \ff(\xx)$ with  initial condition $\xx(t_0)=\widetilde{\xx}_0$, and $\xx_0$   satisfies $\|\xx_0 - \widetilde{\xx}_0\| \le \euler_1$.
For any $\euler>\euler_1>0$, there exists $h>0$ s.t.
\begin{equation*}
\begin{array}{lll}
\label{eq:odehcsp}
   \dot{\xx} = \ff(\xx),\  \xx(t_0)=\widetilde{\xx}_0. \mbox{$\qquad$ \textbf{and}} \\
    (\pwait\ h; \xx:= \xx+h\PHI(\xx,h))^N; \pwait\ \hP; 
 \xx:= \xx+\hP \PHI(\xx,\hP)
 \end{array}
\end{equation*}
are $(h,\euler)$-approximately bisimilar on $[t_0,t_0+T_o]$.
\label{theorem:ODEDiscrete}
\end{theorem}

\subsection{Source Language HCSP}\label{Source}
The syntax for HCSP  is given as follows.
\[  
\begin{array}{lll}  
%e & ::= & c \mid x \mid a[k] \mid f(e) \\
	p  & ::= & 
	\pskip \mid
	x := e \mid 
	ch?x \mid 
	ch!e \mid 	
	p_1; p_2 \mid 
   B \rightarrow p \mid  
	p_1 \sqcup p_2 
 \\
	& & \mid
	 p^* \mid 
	\evo{x}{e}{B} \mid  \exempt{\evo{x}{e}{B}}{i\in I}{ch_i*}{p_i} \\
	pc &::=& p \mid pc_1\|_{cs} pc_2 %\vspace*{-1mm}
\end{array} 
\]
where $e$ represents expressions, $p$ a sequential HCSP process, and $pc$ the parallel composition of processes.  $x$ denotes variables,  $B$ Boolean expressions, $ch, ch_i$ channel names. 
The meanings of $\pskip$, assign, sequential,  conditional, internal choice and repetition are as usual. We explain the intuitive meaning of the additional constructs as follows:
\begin{itemize}
	\item The input $ch?x$ receives a value along the channel $ch$ and assigns it to variable $x$; and the output $ch!e$ sends the value of $e$ along $ch$. Each of them may block waiting for the corresponding dual party to be ready.
	\item The repetition \sm{p^*} executes $p$ for a nondeterministic finite number of times.
    \item \sm{\evo{x}{e}{B}} is the continuous evolution, which evolves continuously according to the ODE \sm{\vec{\dot{x}}=\vec{e}} as long
	as the \emph{domain} $B$ holds, and terminates when $B$ becomes false. 
 %In order to guarantee the existence and uniqueness of the solution of any ODE,   
%the right side  $\vec{e}$ is required to 
% satisfy the \emph{local Lipschitz condition}. 
	%For easing our discussion later, we assume $B$ to be open in what follows. But such assumption can be dropped as any closed set can be  represented as the  interaction of infinite many open sets. 
	\item Interrupt \sm{\exempt{\evo{x}{e}{B}}{i\in I}{ch_i*}{c_i}} behaves like \sm{\evo{x}{e}{B}}, except it is preempted as soon as one of the communication events \sm{ch_i*} takes place, and then is followed by the corresponding \sm{c_i}. %Notice that, if the continuous evolution terminates (reaches the boundary of $B$) before a communication in \sm{\{ch_i*\}_{i\in I}} occurs, the process terminates immediately. 
	\item \sm{pc_1\|_{cs} pc_2} behaves as $pc_1$ and $pc_2$ run independently except that all communications along  common channels $cs$ are synchronized  between $pc_1$ and $pc_2$. 
 \newcomment{We assume that variables of $pc_1$ and $pc_2$ are disjoint, and no same channel direction (e.g. $ch!$) occurs in both $pc_1$ and $pc_2$.}
\end{itemize}
Some other constructs  can be defined as derived. For example, $\pwait\ d$ is an abbreviation for $t:=0;\langle \dot{t}=1\&t<d \rangle$. 

\begin{figure*} \centering  
{\small 
 \begin{eqnarray*} 
&\prftree[r]{}{(x:=e, s) \xrightarrow{\tau}(\pskip, s[x \mapsto e])}
\quad 
\prftree[r]{}{(c_1, s) \xrightarrow{e}(c'_1, s')}{(c_1 ; c_2, s) \xrightarrow{e}(c'_1 ; c_2, s')}
\quad
\prftree[r]{}{(\pskip; c, s) \xrightarrow{\tau}(c, s)}
\quad
\prftree[r]{}{s(B)}{\left(B \rightarrow c, s\right) \xrightarrow{\tau}(c, s)}
\\
&\prftree[r]{}{\neg s(B)}{\left(B \rightarrow c, s\right) \xrightarrow{\tau}(\pskip, s)} \quad
\prftree[r]{}{(c_1 \sqcup c_2, s) \xrightarrow{\tau}(c_1, s)}
\quad
\prftree[r]{}{(c_1 \sqcup c_2, s) \xrightarrow{\tau}(c_2, s)}
\quad
\prftree[r]{}{(c^*, s) \xrightarrow{\tau} (\pskip, s)}
\quad
\prftree[r]{}{(c, s) \xrightarrow{e}(c', s')}{(c^*, s) \xrightarrow{e}(c' ; c^*, s')}
\\
&
\prftree[r]{}{(ch!e, s) \xrightarrow{\langle ch!,s(e)\rangle} (\pskip, s)}
\quad
\prftree[r]{}{(ch!e, s) \xrightarrow{\langle d, \{ch!\}\rangle}  (ch!e, s)}
\quad
\prftree[r]{}{(ch!e, s) \xrightarrow{\langle \infty, \{ch!\}\rangle} (\pskip, s)} 
\\
& \prftree[r]{}{(ch?x, s) \xrightarrow{\langle ch?, v\rangle} (\pskip, s[x \mapsto v])}
\quad 
\prftree[r]{}{(ch?x, s) \xrightarrow{\langle d, \{ch?\}\rangle} (ch?x, s)}
\quad
\prftree[r]{}{(ch?x, s) \xrightarrow{\langle \infty, \{ch?\}\rangle} (\pskip, s)} 
\\
& 	 \prftree[r]{}
		{\begin{array}{cc}
				\vec{p} \mbox{ is a solution of the ODE $\vec{\dot{x}}=\vec{e}$} \\
			\vec{p} (0) = s(\vec{x}) \quad \forall t\in[0,d).\,s[\vec{x}\mapsto \vec{p}(t)](B)
		\end{array}}
		{(\evo{\vec{x}}{\vec{e}}{B}, s) \xrightarrow{\langle d, \{\} \rangle}
			(\evo{\vec{x}}{\vec{e}}{B}, s[\vec{x}\mapsto \vec{p}(d)])}
		\qquad
		\prftree[r]{}
		{\neg s(B)}
		{(\evo{\vec{x}}{\vec{e}}{B}, s) \xrightarrow{\tau} (\pskip, s)} & \\[2mm] 
  &  \prftree[r]{}
		{\begin{array}{cc}
				\vec{p} \mbox{ is a solution of the ODE $\vec{\dot{x}}=\vec{e}$} \quad
				\vec{p}(0) = s(\vec{x}) \quad \forall t\in[0,d).\,s[\vec{x}\mapsto \vec{p}(t)](B)
		\end{array}}
		{(\exempt{\evo{x}{e}{B}}{i\in I}{ch_i*}{c_i}, s)
		\xrightarrow{\langle d, \rdy(\cup_{i\in I} ch_i*) \rangle} 
			(\exempt{\evo{x}{e}{B}}{i\in I}{ch_i*}{c_i}, s[\vec{x} \mapsto \vec{p}(d)]}
		& \\[0.3mm]
 	&  \prftree[r]{}
		{\neg s(B)}
		{(\exempt{\evo{x}{e}{B}}{i\in I}{ch_i*}{c_i}, s) \xrightarrow{\tau}
			(\pskip, s)}
			\quad
		\prftree[r]{}
		{i\in I}{ch_i* = ch!e}
		{(\exempt{\evo{x}{e}{B}}{i\in I}{ch_i*}{c_i}, s)
		\xrightarrow{\langle ch!,s(e)\rangle}  {(Q_i, s)}}
		& \\[0.3mm]
 	& 
		\prftree[r]{}
		{i\in I}{ch_i* = ch?x}
		{(\exempt{\evo{x}{e}{B}}{i\in I}{ch_i*}{c_i}, s) \xrightarrow{\langle ch?,v\rangle} {(c_i, s[x\mapsto v])}}
\\
 &\prftree[r]{G-Tau }
{i\in I \quad (p_i, s_i) \xrightarrow{\tau} (p_i',s_i') }
{(pc, s) \xrightarrow{\tau}_h (pc[p_i'/p_i], s[s_i'/s_i])} \quad
\prftree[r]{G-Comm}
{i, j \in I \quad i \neq j \quad (p_i,s_i) \xrightarrow{\langle ch!, v\rangle} (p_i',s_i')}
{(p_j,s_j) \xrightarrow{\langle ch?, v\rangle}
(p_j',s_j')}
{(pc, s) \xrightarrow{\langle ch,v\rangle}_h
	(pc[p_i'/p_i, p_j'/p_j], s[s_i'/s_i, s_j'/s_j])} & 
\\
&\prftree[r]{G-delay}
{\forall i \in I. (p_i,s_i) \xrightarrow{\langle d, \rdy_i\rangle} (p_i',s_i')}{\forall i, j \in I. i\neq j \Rightarrow \compat(\rdy_i, \rdy_j)}
{(\|_{i\in I}p_i, s) \xrightarrow{\langle d,\cup_{i\in I}\rdy_i\rangle}_h
	(\|_{i\in I}p_i', \uplus_{i\in I}s_i')} & 
\end{eqnarray*}
}  
\caption{Small-step operational semantics of HCSP}
\label{fig:fullhcspsemantics}  
\end{figure*}  
 
Fig.~\ref{fig:fullhcspsemantics} presents the small-step semantics for HCSP.  Two types of transitions are introduced: 
$(p, s) \xrightarrow{e} (p', s')$ defines that a sequential HCSP process $p$ executes from initial state $s$ in one step, produces event $e$ and results in statement $p'$ and state $s'$; and $(pc, s) \xrightarrow{e}_h (pc', s')$ defines one step execution of a parallel HCSP process.  
Here states $s, s' \in \textit{Vars} \rightarrow\textit{Values}$ assign values to variables of $p$. 
%An \emph{event} $e$  defines one step execution of observable behavior for a HCSP process. 
$\tau$ represents an internal discrete event. A \emph{communication event} $\langle ch\triangleright, v\rangle$, where $\triangleright$ is one of $?$, $!$, or nothing, indicating input, output, and synchronized input/output (IO) event, respectively, where $v$ is the transferred value. 
A \emph{wait event} $\langle d, \rdy\rangle$, represents an evolution of time length $d > 0$ with a set of ready channels that are waiting for communication during this period.
We denote the set of the above events by $\textit{HEvts}$. 
A \emph{ready set} is a set of channel directions, indicating that these channel directions are waiting for communication. \newcomment{Two ready sets \sm{\rdy_1} and \sm{\rdy_2} are \emph{compatible}, denoted by \sm{\compat(\rdy_1, \rdy_2)}, if there does not exist a channel $ch$ such that \sm{ch?\in \rdy_1 \wedge ch!\in \rdy_2} or \sm{ch!\in \rdy_1 \wedge ch?\in \rdy_2}. }

We explain some of the transition rules below: 
\begin{itemize}
   \item
  For output  $ch!e$, there are three cases depending on whether the communication occurs immediately: it may occur immediately, wait for some finite time or wait  indefinitely (producing a wait event).  
 
 \item  The execution of $\evo{\vec{x}}{\vec{e}}{B}$ produces an execution duration of the ODE with initial state, represented as a wait event. $B$ must become false at the end, while remaining true before that. During the evolution, the ready set is empty.

 \item For interruption $\exempt{\evo{x}{e}{B}}{i\in I}{ch_i*}{p_i}$, all the communication directions in $\{ch_i*\}_{i\in I}$ become ready at the very beginning. After that, communications have a chance to interrupt up to the time at which the ODE reaches the boundary. 

 \item \newcomment{For parallel composition, without loss of generality, suppose $pc$ is a parallel composition of sequential processes $\{p_i\}_{i\in I}$, and $s$ is a disjoint union of the states for all $p_i$s, i.e. $s = \bigcup_{i\in I}s_i$.  $pc[p_i'/p_i]$ returns a new process by substituting $p_i'$  for $p_i$ in $pc$, and $s[s_i'/s_i]$ the same. }
There are three cases: if one process among $pc$ performs a $\tau$ step, then $pc$ can perform the same $\tau$ step (G-Tau); if two processes synchronise over a same channel, then $pc$ performs a communication immediately (G-comm); if all processes of $pc$ can perform a wait  duration $d$ and their ready sets are mutually compatible, then $pc$ performs a wait duration  $d$, joining all  ready sets together (G-delay). 

 \item The semantics of other compound constructs is defined by structural induction. 
 \end{itemize}

%The semantics of a whole HCSP process $pc$
%is defined inductively from its components. 

%\end{itemize}

\section{Target Language C: Syntax and Semantics}\label{Target}
\begin{figure*}[h]
 \centering 
\[
\begin{array}{llll}
 \mbox{Expressions } & e  &::=& d \mid x \mid \ereturn \mid \vreturn \mid a[k] \mid   e \textbf{ op } e \mid ... \\
\mbox{Statements }  &  c  &::=& \pskip \mid  x = e  \mid x=f(\overline{e}) \mid   g(\overline{e}) \mid c_1; c_2 \mid \IFE{B}{c_1}{c_2} \\
&&& \mid \cwhile{B}{c} \mid \for{c_1}{e}{c_2}{c_3}\mid  \creturn{e} \mid \create{tid}{F}{\overline{e}}\\
   &&& \mid \lock{l} \mid \unlock{l} \mid \wait{cv}{l}  \mid \signal{cv} \mid \join{tid}  \mid x \leftarrow \ereturn.Es \\
\mbox{Variable Decls }  & decl &:=& \overline{T\ x} \quad \mbox{Function Decls } 
   \textit{F}  ::=  T_1\ f(decl_1) \{decl_2; c\} \\
\mbox{Programs }& P &::=& decl; \overline{F}; \texttt{main}   
\end{array}
\]
\caption{Syntax of Subset of Multi-threaded C}  
\label{fig:syntaxofC}
\end{figure*}

We consider a subset of concurrent C with POSIX threads as the target language of code generation.  The abstract syntax is defined in Fig.~\ref{fig:syntaxofC}. Here $d$ denotes constants, $x$ variables, $\ereturn$ and $\vreturn$ introduced for semantic use to record the return value of a function and check whether a return occurs so that the remaining code of the function will not execute,  $a[k]$ array elements, $\textbf{op}$ arithmetic or Boolean operators,  $B$ Boolean expressions, $tid$ the ID of a thread, $l$ mutex, $cv$ condition variables, $f, g$ function names, $\overline{T\ x}$ an abbreviation for a sequence of variable declarations with the form $T_1\ x_1; \cdots; T_n\ x_n$.

The remaining statements define the thread APIs for achieving concurrency, provided by the POSIX thread library of C: 
%The meanings are explained below.
\begin{itemize}
    \item $\create{tid}{F}{\overline{e}}$ spawns a new thread $tid$, and starts execution by invoking function $F$ with arguments $\overline{e}$. 
    
    \item $\clock{l}$ locks mutex $l$ and gains exclusive access to the data protected by $l$. 
    
    \item $\cunlock{l}$ releases mutex $l$ and in consequence another thread is allowed to acquire $l$ and use the shared data. 

    \item $\cwait{cv}{l}$  blocks on condition variable $cv$ and automatically releases mutex $l$. As soon as $cv$ is signaled by another thread, \newcomment{it is unblocked on this signal and turns to re-acquire mutex $l$.} 

    \item $\signal{cv}$ signals the condition variable $cv$ to the thread that is blocked on it and in consequence the thread is released to execute. 

    \item $\join{tid}$ waits for the thread $tid$ to terminate. 
\end{itemize}
The last $x\leftarrow \ereturn.Es$ represents that $x$ is set to be the return value $\ereturn$ in local states $Es$. Same as $\ereturn$ and $\vreturn$, it is introduced  for defining the small-step semantics of function calls with return values. 
A function declaration $F$ includes a return type $T_1$,   a function name $f$, and a body consisting of a sequence of local variable declarations and a command $c$. 
At the end, a C program $P$ is composed of a sequence of global variable declarations $decl$, a sequence of function declarations $\overline{F}$, and the \texttt{main} function as the entry point of the
program.

\subsection{Small-step Semantics}

\subsubsection{Notations}

The semantics of C statements is defined by two judgements. They are parameterized by a static environment $\Gamma$,  a thread pool $T$, a global state $G$, and a local state set $Es$. $\Gamma$ maps a function name to its declaration, $T$ maps each active thread to its code, $G$  maps global variables to their values,  and $Es$ maps each thread to its local state, which in turn maps local variables of the thread to values. Below list the judgements defining the one step execution of a thread and of a thread set consisting of multiple threads resp.:  
\begin{itemize}
    \item $\Gamma, tid \vdash (c, T, G, Es) \rightarrow_c (c', T', G', Es')$, stating that under static environment $\Gamma$ and thread $tid$, statement $c$ executes to $c'$ in one step,  changing thread pool $T$ (due to thread creation), global state $G$, local states $Es$ to $T'$, $G'$ and $Es'$ resp. 

    \item $\Gamma \vdash (T, G, Es) \rightarrow (T', G', Es')$, stating that under static environment $\Gamma$,  thread pool $T$ executes, leading to the continuation $T'$, changing global state $G$ and local states $Es$  to $G'$ and $Es'$ resp.

\end{itemize}
%Events include $\elock$, $\eunlock$, corresponding to the lock and unlock of a mutex; $\ewait$, meaning the thread needs to wait for sometime to execute; $\esignal$, representing a thread issues a signal.  We will use $\textit{CEvts}$ to denote these events. 

\begin{figure*} \centering  
{\small 
 \begin{eqnarray*} 
 &
  \prftree[r]{Assign}{
  \begin{array}{cc}
     Es(tid)(\vreturn) = 0 \quad \seman{e}_{G, Es} = v \quad G'= (v \in \textbf{GV})?G[x \mapsto v]:G\\ 
       Es'= (v \in \textbf{GV})?Es: Es[tid \mapsto Es(tid)[x \mapsto v]] 
  \end{array}}{\Gamma, tid \vdash (x=e, T, G, Es) \rightarrow_c  (\epsilon, T, G', Es')} \\
  &
  \prftree[r]{Funcr}{Es(tid)(\vreturn) = 0 \quad \Gamma(f) = T\ f(\overline{T\ x})\{\overline{U\ y}; c\}\quad \seman{\overline{e}}_{G, Es} = \overline{v} \quad E=\{\overline{x} \mapsto \overline{v}, \overline{y} \mapsto \overline{\textbf{0}}\}}{\Gamma, tid \vdash (x=f(\overline{e}), T, G, Es) \rightarrow_c  (c;x\leftarrow\ereturn.Es, T, G, [tid \mapsto E, \vreturn\mapsto 0])} \\
    &
  \prftree[r]{FunEnd}{Es(tid)(\vreturn) = 1}{\Gamma, tid \vdash (x\leftarrow \ereturn.Es', T, G, Es) \rightarrow_c  (\epsilon, T, G, Es'[x \mapsto Es(\ereturn), \vreturn \mapsto 0])} \\
  &
  \prftree[r]{Ret}{Es(tid)(\vreturn) = 0}{\Gamma, tid \vdash (\creturn{e}, T, G, Es) \rightarrow_c  (\epsilon, T, G, Es[tid \mapsto Es(tid)[\ereturn \mapsto \seman{e}_{G, Es}], \vreturn \mapsto 1])} \\
 &
  \prftree[r]{SeqT}{\Gamma, tid \vdash (c_1, T, G, Es) \rightarrow_c (c_1', T', G', Es') }{\Gamma, tid \vdash (c_1;c_2, T,  G, Es) \rightarrow_c  (c_1';c_2,   T', G', Es')} \\
&\prftree[r]{SeqF}{}{\Gamma, tid \vdash (\epsilon;c_2, T,  G,  Es) \rightarrow_c (c_2, T, G, Es)}\\
  &
  \prftree[r]{WhileT}{Es(tid)(\vreturn) = 0 \quad \seman{B}_{G, Es} = \textbf{True} \quad \Gamma, tid \vdash (c, T, G, Es) \rightarrow_c (c', T', G', Es') }{\Gamma, tid \vdash (\cwhile{B}{c}, T,  G, Es) \rightarrow_c  (c'; \cwhile{B}{c},   T', G', Es')} \\
&\prftree[r]{WhileF}{Es(tid)(\vreturn) = 0 \quad \seman{B}_{G, Es} = False }{\Gamma, tid \vdash (\cwhile{B}{c},T,  G, Es) \rightarrow_c (\epsilon, T, G, Es)}
   \\ 
 & \prftree[r]{Ret'}{\neg (c \equiv x\leftarrow \ereturn.Es \vee c\equiv c_1;c_2) \quad  
  Es(tid)(\vreturn) = 1}{\Gamma, tid \vdash (c, T, G, Es) \rightarrow_c  (\epsilon, T, G, Es)} \\
 &\prftree[r]{Create}{Es(\vreturn) = 0}{\seman{\overline{e}}_{G, Es} = \overline{v} }{\Gamma, tid \vdash (\create{tid'}{F}{\overline{e}}, T, G, Es) \rightarrow_c (\epsilon, T[tid'\mapsto F(\overline{v})], G, Es[tid'\mapsto Es(tid)]}
   \\[2mm]  
%   & \prftree[r]{LockF}{\exists tid' \in \textit{Threads}. tid' \neq tid \wedge G(l) = tid' }{\Gamma, tid \vdash (\lock{l}, T, G, Es) \xrightarrow{\ewait }_c (\lock{l}, T, G, Es)} \\
  & \prftree[r]{Lock}{Es(\vreturn) = 0}{G(l) = \bot }{\Gamma, tid \vdash (\lock{l}, T, G, Es) \xrightarrow{ }_c (\epsilon, T, G[l \mapsto tid], Es)} \ \prftree[r]{Unlock}{Es(\vreturn) = 0}{G(l) = tid}{\Gamma, tid \vdash (\unlock{l}, T, G, Es) \xrightarrow{ }_c (\epsilon, T, G[l\mapsto \bot], Es)} & \\[2mm]
 % & \prftree[r]{WaitN}{\sigma(t)(cv) = 1 \quad \sigma(t)(l) = 1}{(\cwait{t}{cv}{l},  \sigma) \xrightarrow{ } (\epsilon,  \sigma)} \quad\quad
 & \prftree[r]{WaitF}{Es(\vreturn) = 0}{G(cv) = 0 \quad G(l) = tid}{\Gamma, tid \vdash(\wait{cv}{l}, T,  G, Es) \xrightarrow{ }_c (\cwait{cv}{l}, T, G[l \mapsto \bot], Es)} \quad &\\[2mm] 
% & \prftree[r]{WaitS}{G(cv) = 0 \quad G(l) \neq tid}{\Gamma, tid \vdash(\wait{cv}{l}, T,  G, Es) \xrightarrow{\ewait }_c (\cwait{cv}{l}, T, G, Es)} \quad &\\[2mm] 
  &  \prftree[r]{WaitT}{Es(\vreturn) = 0}{G(cv) = 1 \quad  
 G(l) \neq tid}{\Gamma, tid \vdash (\cwait{cv}{l}, T, G, Es) \xrightarrow{ }_c (\lock{l}, T, G[cv \mapsto 0], Es)}\\
&\prftree[r]{Signal}{Es(\vreturn) = 0}{\Gamma, tid \vdash (\signal{cv}, T,  G, Es) \xrightarrow{  }_c (\epsilon,  T, G[cv \mapsto 1], Es)}  \ 
\prftree[r]{Join}{Es(\vreturn) = 0}
{T(tid') = \epsilon}
{\Gamma, tid \vdash (\join{tid'}, T, G, Es) \xrightarrow{ }_c  (\epsilon, T, G, Es)}\\
 &  
\prftree[r]{Threads}
{
\begin{array}{cc}
     tid \in   dom(T) \quad T(tid) = c \quad Es(tid) = E \quad
    \Gamma,  tid \vdash (c, T, G, Es) \rightarrow_c (c', T', G', Es')
\end{array} }
{\Gamma \vdash (T, G, Es) \rightarrow_t (T'[tid \mapsto c'], G', Es')}
\\
% & \prftree[r]{Ret'}{\neg (c \equiv x\leftarrow \ereturn.Es \vee c\equiv c_1;c_2) \quad  
%   Es(tid)(\vreturn) = 1}{\Gamma, tid \vdash (c, T, G, Es) \rightarrow_c  (\epsilon, T, G, Es)} \\
%& 
%\prftree[r]{Env1}
%{Es'(tid) = Es(tid) }
%{\Gamma, tid \vdash (c, T, G, Es) \stackrel{a}{\dashrightarrow_c}   (c, T', G', Es')}\\
\end{eqnarray*}
}
\caption{Small-step operational semantics of C}
\label{fig:fullcsemantics}  
\end{figure*}

\subsubsection{Semantics}
We present the small-step semantics of multi-threaded C in Fig.~\ref{fig:fullcsemantics}. 
Rule (Assign) changes the value of $x$ to $e$ in $G$ or $Es$ depending on whether it is global variable (denoted by $x \in \textbf{GV}$) or not. Rules (Funcr, Ret, FunEnd) define the semantics for a function call with return value: (Funcr) first looks up the definition of function $f$ in $\Gamma$, then builds a new local state mapping formal parameters $\overline{x}$ to  values of $\overline{e}$ and local variables of $f$ to be their default values (represented by $\textbf{0}$ here), and executes the body statement $c$ from this new state. At the same time, at the end of $c$, $x\leftarrow \ereturn.Es$ is added, to be explained in (FunEnd); (Ret) sets $\ereturn$ to be the value of $e$   and $\vreturn$ to be $1$ indicating that a return value is obtained and the rest code inside the function body will not be executed any more. (FunEnd) executes when $\vreturn$ is 1: it resumes the previous state $Es'$, assigns $x$ by the value of $\ereturn$ recorded in $Es$, and meanwhile, reset $\vreturn$  to 0 indicating that this function body is jumped and thus the following code can execute as normal. 
In conclusion, for handling function call:
\newcomment{Each statement is guarded by $\vreturn=0$, to mean that no return occurs thus the continuation  executes; For the contrary case when $\vreturn = 1$, the continuation will be dropped, indicated by rule (Ret'), until the function call ends, indicated by rule (FunEnd), which transfers the return value back and resets $\vreturn$ to  0 again. }

Rule (Create) spawns a new thread $tid'$, recording the code of $tid'$ in the thread pool to be $F(\overline{v})$, and meanwhile setting its initial local state as the one of parent thread $tid$. 
Rule (Lock) defines that if mutex $l$ is available, then it can be obtained by thread $tid$, by mapping $l$ to be the holding thread $tid$ in the global state $G$. Rule (Unlock) defines that $tid$ releases $l$. The semantics of $\textbf{cwait}$ is specified by two rules: At first, $tid$ locks $l$ and $cv$ is false,  $t$ releases $l$ (WaitF); then as soon as $cv$ is signaled thus becomes true, $\textbf{cwait}$ stops waiting,   and then it needs to acquire mutex $l$ again and thus is equivalent to executing $\textbf{lock}$, and at the same time resetting $cv$ to be false (WaitT).  $\signal{cv}$ signals $cv$ to some thread who is waiting on it (Signal). 
Rule (Join) defines that when thread $tid'$ completes the execution of its code, $\join{tid'}$ terminates directly. 
Rule (Threads) defines the execution of multiple threads in $T$, which randomly selects an available thread $tid$ to execute and updates the thread pool, the global and local states correspondingly. 
%Obviously, when $T$ only contains one thread, it degenerates to the case for the execution of one single thread; when $T$ contains all active threads, it defines the  execution of the entire program. 
%\subsubsection{Big-step Semantics}
%Above we define the small-step semantics of statements of multi-thread C, based on which the behavior of parallel composition of multiple threads is defined as the interleaving executions of them. The big-step semantics of a thread set, denoted by $\Gamma \vdash (T, G, Es) \Rightarrow_t (T', G', Es')$, can be defined based on the transition closure of the small-step semantics, i.e. $\Rightarrow_t = \rightarrow_t^*$.  
%\subsubsection{Semantics of C Programs}
%The semantics of a whole C program can be defined based on the transition rules of statements, after instantiating $\Gamma$, $T$, $G$ and $Es$ resp.
 
Above we have defined the semantic rules for both HCSP and C subset, based on which the transition systems for them can be built.   

\section{From HCSP to C}\label{HCSP2C}

\subsection{Auxiliary Variables and Functions}
In order to transform HCSP to C, some auxiliary variables and functions are introduced in order to achieve synchronization between multiple threads. 

\subsubsection{Global and Local Clocks}
In order to synchronize the executions of threads in parallel, we introduce a local clock for each thread $i$, denoted by $\texttt{localTime[i]}$, to record the local execution time of $i$. If a thread is waiting with a time limit (and possibly for some communications), its $\texttt{localTime}$ is set to that time limit. If the thread is only waiting for communication, its $\texttt{localTime}$ is set to infinity (with $\texttt{DBL\_MAX}$ used in practice). A global clock $\texttt{currentTime}$ is used to record the global execution time, and to coordinate the execution of all threads. It equals the minimum of all local clocks, thus every time a local clock makes progress, $\texttt{currentTime}$ will be updated and in consequence threads whose time limit has reached will be woken up and be notified to execute.

\subsubsection{Thread States}

There are six possible forms of thread states that indicate the execution status of a thread:
\begin{itemize}
\item \texttt{Stopped}, representing the thread has reached the end of execution; 

\item \texttt{Waiting}, representing that the thread is waiting for the global clock with a time limit, specified by $\texttt{localTime}$. When the global clock reaches that time limit, the thread is released to run. 

\item \texttt{Available}, representing that the thread is waiting for a communication event. As soon as it receives the signal from another thread denoting that a compatible communication event is ready, it is woken up to run. 

\item \texttt{Waiting\_Available}, representing that the thread is waiting both for the global clock to reach a time limit, and for a communication event. Either of them can release the thread to run.

\item \texttt{Running}, representing that the thread is able to execute, not waiting for the global clock or communication events;

\item A non-negative number $i$, representing that a communication is occurring on channel $i$. This is an intermediate state used to synchronize communication, meaning that the thread will carry out a communication on channel $i$, but has not finished doing so.
\end{itemize}

The thread states are shared by all threads and used for coordinating the execution of them. In our implementation, each thread is denoted by an ID among $\{0, 1,\cdots, N\!-\!1\}$, where $N$ is the number of threads. An array $\texttt{threadState}[N]$ is used to record the execution states of each thread. 

\subsubsection{Channels in C}
In order to realize synchronized communication in C, we introduce global variables related to channels. In HCSP, each channel $ch$ includes two ends: $ch?$ for input and $ch!$ for output, and they synchronize both on time and values transferred along $ch$. In C implementation, we introduce a structure \texttt{Channel} to define each single channel end, which is $(\texttt{type, channelNo, pos})$, representing its type (0 for input and 1 for output), channel ID number (represented by natural numbers 0, 1, ...), and the pointer referring to the value it holds. Especially, each pair of input and output ends of a channel have the same channel ID. We also define three arrays for achieving synchronization between inputs and outputs: for each channel $i$,  
\texttt{channelInput[i]} and \texttt{channelOutput[i]} record the thread that is ready on corresponding input and output alone channel $i$, and thus available to participate in the communication. The default value is $-1$, denoting that no thread is available for the corresponding input or output; \texttt{channelContent} acts as a buffer to save the values transmitted along channels: the output ends write to it while the input ends read from it. 

\subsubsection{Locks and Condition Variables}
A mutex $\texttt{mutex}$ is introduced to protect the shared resources of threads. To access these resources, a thread must acquire $\texttt{mutex}$ first. Moreover, to achieve synchronization between threads, an array $\texttt{cond}$[N] of condition variables are introduced, one for each thread.

\subsection{Transformation of HCSP}

\subsubsection{Continuous evolution}
For continuous evolution $\langle \dot{\xx} = \ff(\xx) \& B \rangle$, we will first discretise it in HCSP and then transform the discretisation to C code.  
Before discretisation, the discretized time step $h$ should be computed first. Given a HCSP process $P$, suppose the upper bound for the execution time of $P$ is $T_o$, and the value precision is $\varepsilon$, then $h$ is computed as follows:
\begin{itemize}
    \item First, collect the set of ODEs of $P$, denoted by $\{ode_1(\overline{x_1}), \cdots, \ ode_k(\overline{x_k})\}$. Among them, for any $i \neq j$,  $\overline{x_i}$ and $\overline{x_j}$ are continuous variables of the two ODEs resp. and they might have common variables;
    \item Next we compute the upper bound of the discretized step. On one hand, suppose the Lipschitz constants for  the $k$ ODEs exist and are $L^1, \cdots, L^k$, and the constants for the Runge-Kutta method are ${C_1}^1, \cdots, {C_1}^k, {C_2}^1, \cdots, {C_2}^k $, resp., as presented in Prop.~\ref{theorem:globalerror}. Then  compute $M(h) \leq \frac{\varepsilon}{2}$ by assigning $L, C_1, C_2$ to be $\max\{L^i\}_{1\leq i \leq k}$, $\max\{{C_1}^i\}_{1\leq i \leq k}$,$\max\{{C_2}^i\}_{1\leq i \leq k} $ resp., which finally obtains an upper bound for the discretized time step, denoted by $h_1$. We must have $h \leq h_1$. 

    \item On the other hand, suppose for each ODE $ode_i(\overline{x_i})$ of the form $\langle \dot{\xx_i} = \ff_i(\xx_i) \& B_i \rangle$, the derivative $\ff_i(\xx_i)$ is bounded over the interval $[0, T_o]$, satisfying $\|\ff_i(\xx_i)\| \leq U_i$. Then the distance between any two states at time $a$ and $b$ within one time step is bounded by $U_i \cdot \|a-b\|$. 
    %As the distance between any two states within one discretized time step must be less than $\frac{\varepsilon}{2}$, we have 
    Let $U_i \cdot h < \frac{\varepsilon}{2}$, then $h < \frac{\varepsilon}{2U_i}$ for any $i$. 

    \item At the end, let $h$ be  $\min (h_1, \min_{1 \leq i \leq k}\{\frac{\varepsilon}{2 U_i}\})$. 
\end{itemize}
% \newpage

After $h$ is calculated, for any continuous evolution $\langle\dot{\xx} = \ff(\xx) \& B\rangle$ occurring in $p$, it  is transformed to the following discrete HCSP process according to Theorem~\ref{theorem:ODEDiscrete}:
\begin{align}
\label{eqn:ODE}
&j := 1; 
     (\neg (N(B, \varepsilon) \wedge N^n(B, \varepsilon)) \rightarrow j = 0;\notag\\
& \qquad \qquad     j=1 \rightarrow   \pwait\ h; \xx:= \xx+h\PHI(\xx,h))^N; \\
&   N(B, \varepsilon) \wedge N^n(B, \varepsilon) \rightarrow \texttt{stop}; \notag
\end{align} 
where $N = \lceil \frac{T-T_0}{h}\rceil$. \newcomment{Given a Boolean formula $B$, which can be considered as a set of states satisfying $B$,  $N(B, \varepsilon)$ denotes the $\varepsilon$-neighbourhood of $B$, representing the set $\{a\,|\,\exists b. |a-b|<\varepsilon \wedge b \in B\}$ (obviously $B \subset N(B, \varepsilon)$);} and $N^n(B, \varepsilon)$ is an abbreviation of $N(B[\xx \mapsto \xx + h\PHI], \varepsilon)$, i.e. the $\varepsilon$-neighbourhood of $B$ at next discretized time step. Instead of $B$, $N(B, \varepsilon) \wedge N^n(B, \varepsilon)$ is used to judge whether to continue evolving according to the ODE. 
\oomit{If $B$ holds for the original ODE at time $t$, then $N(B, \varepsilon)$ must hold for the discretized values at $t$, but $N^n(B, \varepsilon)$ may not hold. 
But for the latter case, $B$ must be false at next time step.
If $B$ does not hold at time $t$, $N^n(B, \varepsilon)$ must not hold at next time step due to the definition of robustly-safety of $p$ (to be given later).} \newcomment{With the help of robustly safe condition (to be explained in next section), the discretized implementation can always be guaranteed to escape within a tolerance. For the example shown in Fig.~\ref{fig:Robustlysafe}, the discretization escapes at next step $(n+1)h$ when the ODE violates $B$ between $nh$ and $(n+1)h$. With the C implementation of $\pwait\ h$ by time delay, the C code is a direct translation of (\ref{eqn:ODE}). }

 \oomit{
Below give the corresponding C code, where \texttt{Phi} and \texttt{eps} correspond to $\PHI$  and precision $\varepsilon$ resp.
\begin{lstlisting}[style=mystyle,language=custom]
void odec(int tid) {
    int i = 1;
    while(i==1){
      if(!N(B, eps) | !N^n(B, eps)) i = 0;
      if (i==1) {delay(h); x = x + h*Phi(x,h);} 
      if (currentTime >= T) {i = 0; stop;}
   } }
\end{lstlisting}
}
%The whole structure is a while loop: In each step, the time progresses for $h$ time units  and $x$ are updated  according to Theorem~\ref{theorem:ODEDiscrete} (lines 4-5); then if the enlarged $B$ with $\varepsilon$ precision does not hold for $x$ in this step, or next step, then the loop exits (line 6-7); otherwise, continue and check if the global time reaches the time bound $T$, if yes, all the variables are kept unchanged from this time point and the loop exits. 

\subsubsection{Continuous Interrupt}

Continuous interruption $\exempt{\evo{x}{e}{B}}{i\in I}{ch_i*}{p_i}$ is first transformed to a sequential composition of discrete processes in  HCSP, among which $n, N, \PHI, \varepsilon$ are as defined before. For any HCSP process $p$, $\DHCSP(p)$  transforms $p$ to its discretised version. 
\vspace{-5mm}

{\small
 \begin{align} \label{eqn:odeI}
 &      j_1:=1; j_2 := -1; \notag \\
 &(\neg(N(B, \varepsilon) \wedge N^n(B, \varepsilon)) \rightarrow j_1:=0;\notag \\ 
 &\ j_1 = 1 \wedge j_2=-1 \rightarrow c :=0; \\ 
 &    \qquad \exempt{\langle \dot{c} =1 \& c \leq h\rangle}{i\in I}{ch_i*}{j_2:=i};\notag\\
&  \ j_1 = 1 \wedge j_2 =-1 \rightarrow \xx:= \xx+h\PHI(\xx,h))^N; \notag\\
 %&  \ j_1 = 0 \wedge    j_2 = -1  \rightarrow
%     \\
%&     \qquad \exempt{\pwait\ 0}{i\in I}{ch_i*}{j:=i};\notag\\
&     j_2 \geq 0 \rightarrow \xx:= \xx+c\PHI(\xx,c);  \DHCSP(p_{j_2});\notag\\
&   j_1=1 \wedge  j_2 =-1 \rightarrow \texttt{stop};  \notag  
\end{align}
}

\noindent $c:=0; \langle \dot{c} =1 \& c \leq h\rangle$ is equivalent to $\pwait\ h$. We use it here in order to record the communication interrupt time by variable $c$, used to compute the state  later (guarded by $j_2\geq 0$). The C code corresponding to $\exempt{\pwait\ h}{i\in I}{ch_i*}{j_2:=i}$ is denoted by $\mathsf{wait\_comm}(h, \{ch_i*\}_{i\in I})$, to be explained next. 
%Then the above discrete process can be transformed to C by structural induction.

\begin{figure}[h]
\centering
\includegraphics[scale=0.4]{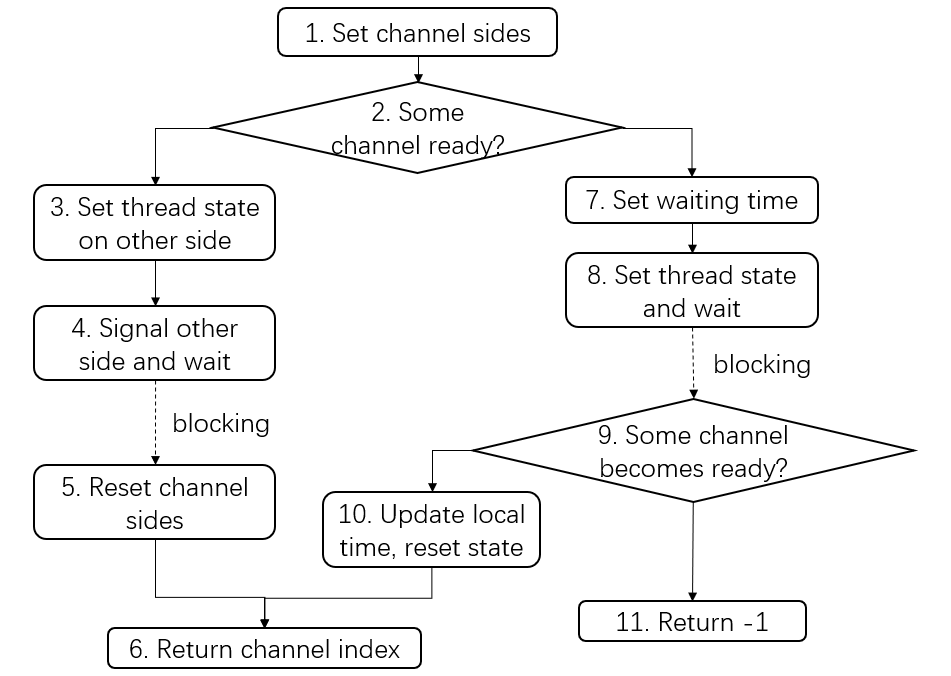}
\caption{The control flow of C code for $\mathsf{wait\_comm}$}
\vspace{-3mm}
\label{fig:waitcomm}
\end{figure}

\subsubsection{Wait communication}
Fig.~\ref{fig:waitcomm} presents the control flow for the C code corresponding to $\mathsf{wait\_comm}(h, \{ch_i*\}_{i\in I})$, after locking the mutex in the beginning: 1, Set \texttt{channelInput} and \texttt{channelOutput} for each communication to the current thread, indicating that $\{ch_i*\}_{i\in I}$ are all available, and meanwhile write values to the channel content buffer from outputs;
2, Look through the set of compatible channel ends  to see if any is ready; 3,
If some channel end is ready, set the state of the other thread to that channel  (reserving communication on the other side); 4, 
Signal to the other thread and wait; 5,
On returning from wait, reset thread state to be \texttt{RUNNING} and all channel sides corresponding to $\{ch_i*\}_{i\in I}$ to be unready; 6,
In case a communication is performed, the function returns the channel index for later use; 7, If no channel is ready at the beginning, set target waiting time recorded by \texttt{localtime}; 8, Set thread state to be \texttt{AVAILABLE} and wait; 9, On returning from wait, check the thread state to determine which channel among $\{ch_i\}_i$ is matched to perform the communication; 10, Update local time to be the local time of the other thread, and reset all states as in step 5; 11, If no communication occurs during time $h$, return $-1$ and terminate. 
Input and output can be considered as special cases of $\mathsf{wait\_comm}$. 

\newcomment{Above we present the primitive functions related to communication and ODE used for transformation. By calling these primitives, each type of sequential compound processes of HCSP can be transformed inductively. At last, the top HCSP process, in the form of parallel composition, is transformed to a set of threads corresponding to all its sequential sub-processes.  }

In the appendix, we list the generated C code for some HCSP constructs as representatives. 

\section{Correctness of the Transformation}
\label{sec:correct}

%In the previous sections, we defined the semantics of both the source HCSP language, and the target concurrent C programs. We also described the transformation rules from HCSP to C. 
In this section, we present the correctness proof of the transformation. % The result to be proved is that the HCSP process (denoted by $\mathcal{P}$ below) and its corresponding C code (denoted by $\mathcal{C}$ below) are approximate bisimilar.
The proof uses the following strategy, separating the concerns of discretization and reasoning about concurrency into two steps. First, we show that the HCSP process and its discretization are approximately bisimilar. Second, we show that for the discrete fragment of HCSP (in which the discretization of any HCSP program lies), the translation to concurrent C code is (exactly) bisimilar.

We first introduce the notion of robustly safe HCSP processes (Sect.~\ref{sec:robustly-safe}). Then we state the theorem on approximate bisimulation between HCSP and C (Sect.~\ref{sec:approximate-bisimulation}). The first part of the proof, that HCSP  is approximately bisimilar to its discretization, is given partly in Sect.~\ref{sec:proof-bisimulation-continuous} for the case of continuous evolution and interrupt, while others are similar to~\cite{DBLP:journals/tosem/YanJWWZ20} and omitted here. Finally, we present the proof of exact bisimulation between HCSP and the generated C code in Sect.~\ref{sec:proof-discrete-bisimulation}.

\subsection{Robustly Safe HCSP Processes}
\label{sec:robustly-safe}
\newcomment{
Due to the discretization of the ODE, the behaviors of an HCSP process and its transformed C program are not exactly equivalent but allow a difference within given precisions. An HCSP process must be robust to tolerable errors.  Especially, the discretization should preserve the control behavior in alternation like $B \rightarrow p$, and moreover, it should be able to detect the change of domain boundary in continuous evolution like $\langle \dot{x} = f(x) \& B \rangle$ and locate the changing point within a tolerance (which is the so-called zero-crossing detection and location in hybrid systems~\cite{ZHANG20087967}). Hence, we introduce the notion of \emph{robustly safe processes} with given precisions in Def.~\ref{def:robustsafe}. Let $\phi$ denote a Boolean formula and $\epsilon$ a precision, define $N(\phi, -\epsilon)$ as the set $\{ v\,|\,v \in \phi \wedge \forall u \in \neg \phi.\,|v-u| > \epsilon\}$, which is a subset of $\phi$ and furthermore the distance from all states in it to the boundary of $\phi$ are greater than $\epsilon$.}
\vspace{-2mm}

\begin{figure}[htbp]
  \centering
  \includegraphics[width=0.42\textwidth]{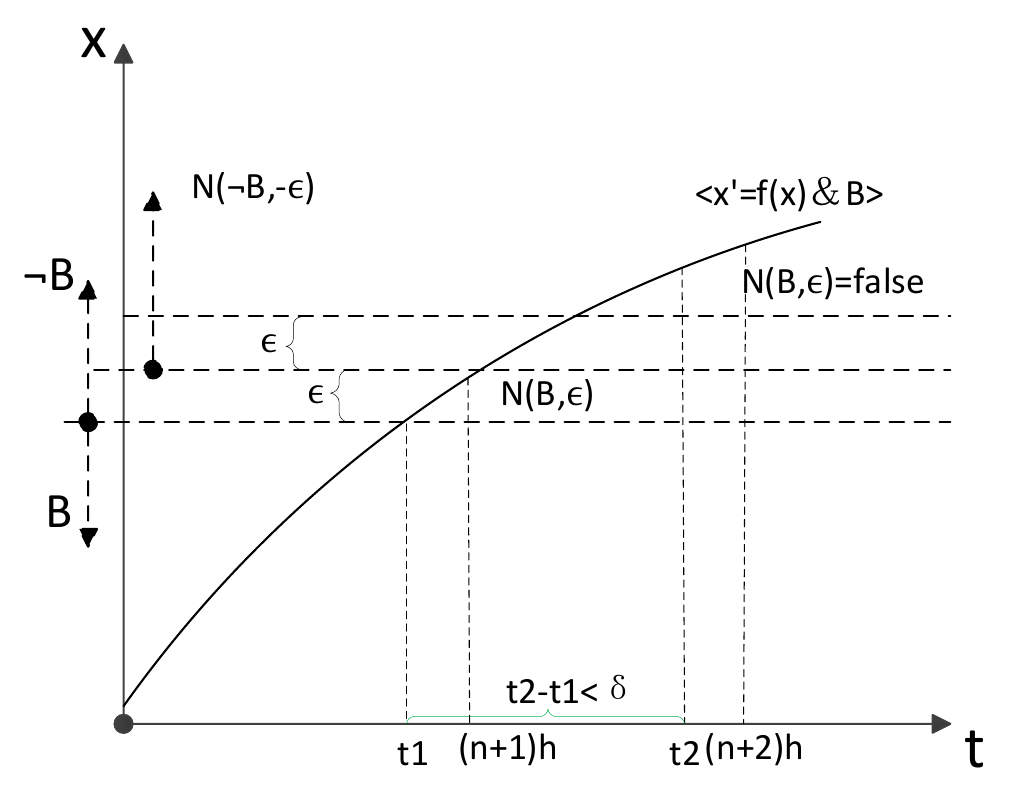}
  \caption{ The $(\delta, \epsilon)$-robustly safe continuous statement.}
  \label{fig:Robustlysafe}
\end{figure}

\begin{definition}[$(\delta, \epsilon)$-robustly safe]
   An HCSP process $P$ is $(\delta, \epsilon)$-robustly safe, for given time precision $\delta>0$ and  value precision $\epsilon>0$, if the following two conditions hold:
   \begin{itemize}
     \item For every alternative process $B \rightarrow Q$ occurring in $P$ with  $B$ depending on continuous variables of $P$, then the states $v$ reached before the execution of $B \rightarrow Q$ satisfy  $v \in N(B, -\epsilon)$ or $v\in N(\neg B, -\epsilon)$; 
    
     \item For every continuous evolution $\evo{\vec{x}}{\vec{e}}{B}$ occurring in $P$, suppose its initial state is  $v_0$, and $B$ turns to false at state $v$ and time $t$,  then there exists $\widehat{t} \in (t, t+\delta)$ s.t. $U(v[\vec{x} \mapsto X(\widehat{t}, \widetilde{\vec{v}}_0)], \epsilon) \subseteq N(\neg B, -\epsilon)$, where $X(t, \widetilde{\vec{v}}_0)$ is the solution of $\dot{\vec{x}} = \vec{e} $  at $t$ with initial value $\widetilde{\vec{v}}_0$.   \end{itemize}
\label{def:robustsafe}
\end{definition}

Def.~\ref{def:robustsafe} requires: (1) All reachable states before $B \rightarrow P$ are  $\epsilon$-far from the boundary of $B$, which is also the boundary of $\neg B$. \newcomment{(2) As shown by Fig.~\ref{fig:Robustlysafe}, when $B$ turns false at time $t_1$, then before time $t_2$ (within $\delta$ tolerance), the ODE continues to go away from the boundary of $B$, at least $2\epsilon$ further at $t_2$. According to our discretization (by checking $N(B, \epsilon) \wedge N^n(B, \epsilon)$), it is guaranteed to   detect the change of $B$ at next step of $t_1$, i.e. $(n+1)h$ in the example.}
%is robust to time precision $\delta$ and value precision $\epsilon$, i.e. the reachable states at escaping point will keep violating $B$ for at least $\delta$ time and furthermore $\epsilon$ far from the boundary of $\neg B$. 
%Thus, as long as the discretized error is less than $\epsilon$, the original HCSP process and the discretized have the same value for $B$ in $B \rightarrow P$, and  escapes $B$ in $\evo{\vec{x}}{\vec{e}}{B}$ within at most $\delta$ time error.
\newcomment{Computing the robustly safe parameters $(\delta, \epsilon)$ is very challenging and some methods for computing them are given in~\cite{DBLP:journals/tosem/YanJWWZ20}.}
%In this paper, we assume the HCSP processes to be transformed are always known to be $(\delta, \epsilon)$ robustly safe for some $\delta$ and $\epsilon $. 

%There are two special cases without requiring robustness conditions for HCSP:  all the control variables (the ones occurring in conditions or domains of ODEs) are discrete, with no approximate error;  or when all ODEs have explicit solutions, so that both the values of continuous variables and the escaping points of ODEs can be calculated precisely. 

\subsection{Approximate Bisimulation between HCSP and C}
\label{sec:approximate-bisimulation}
We then state the main theorem for approximate bisimulation between HCSP and C. \newcomment{Next $\DHCSP_{h, \varepsilon}(P)$ and $\toc_{h, \varepsilon}(P)$ denote the discretized version and the generated C code of HCSP process $P$ with precisions $h$ and $\varepsilon$ resp.}
\begin{theorem}
     Let $P$ be a HCSP process and $T>0$ is an upper bound of time. Suppose $P$ is $(\delta, \epsilon)$-robustly safe, and for any ODE $\dot{x} = f(x)$ occurring in $P$, f is Lipschitz continuous. Then, for any precision  $\varepsilon \in (0, \epsilon]$, there must exist $h>0$ such that $P \cong_{h, \varepsilon} \toc_{h, \varepsilon}(P)$ holds on $[0, T]$. 
     %Especially, if $\delta >0$, $h$ satisfies $\frac{\delta}{2} < h < \delta$. 
     \label{them:weakbisi}
\end{theorem}
 \newcomment{We will prove Theorem~\ref{them:weakbisi} in two steps:   $P \cong_{h, \varepsilon} \DHCSP_{h, \varepsilon}(P)$, and  $\DHCSP_{h, \varepsilon}(P) \cong \toc_{h, \varepsilon}(P)$. }
%The discretized precision $\varepsilon$ must be less than the robustly safe parameter $\epsilon$ and the parameter $\delta$ satisfies 
%$h < \delta < 2h$ to guarantee the escape of the ODE boundary is caught within one time step in discretization (as shown in Fig.~\ref{fig:Robustlysafe}).
%meaning that the discretization can always catch with the original continuous dynamics on the escape of the ODE boundary, after further progressing one more time step (from the exact escaping point) but at most within two steps. 

\subsection{Proof of approximate bisimulation between HCSP and Its Discretization}
\label{sec:proof-bisimulation-continuous}

\oomit{Let 
$\mathcal{TS}_i = \langle S_i, L_i, \rightarrow_i, {S_i}^0, V\rangle, i=1, 2$ be the transition systems of  $\mathcal{P}$ and  $\toc_{h, \varepsilon}(\mathcal{P})$ resp., we need to prove that $\mathcal{TS}_1$ and $\mathcal{TS}_2$ are $(h, \varepsilon)$- approximate bisimilar. We add an intermediate step between HCSP and C for separation of concerns in the proof:  first, prove a HCSP process and its discretisation are approximate bisimilar;  second, prove the discretised HCSP and the C code are bisimilar.   The first part is similar to~\cite{DBLP:journals/tosem/YanJWWZ20}, and omitted here.}

We present the proof for the approximate bisimulation between continuous HCSP processes and the  discrete ones. 

\begin{itemize}  

\item[1)] 
We first show the proof for continuous evolution. Denote the ODE $\langle\xx = \ff(\xx) \& B\rangle$ by $ode_k$, suppose the exact initial value of $\xx$ prior to its execution is $\widetilde{\xx}_0$, and the discretized value is $\xx_0$,  satisfying
$\|\widetilde{\xx}_0 - \xx_0\| \leq  \euler_1 < \varepsilon$.
%where $t_0$ denotes the execution time till the execution of $ode_k$. $M_1(h)$ is as defined in Prop.\ref{theorem:globalerror} by assigning  $L, C_1, C_2$  the upper bounds for the ODEs of $\xx$.  $M_1(h) \leq M(h) < \varepsilon$ must hold. 
Now we consider the transitions of the ODE and the discretised process (\ref{eqn:ODE}).  
\begin{itemize}
    \item If $(\langle\xx = \ff(\xx) \& B\rangle, s_0) \xrightarrow{\langle d, \{\}}(\langle\xx = \ff(\xx) \& B\rangle, s_0[\xx \mapsto \mathbf{p}(d)]$, where $\mathbf{p}$ is the solution of the ODE with initial state $s_0$, then $B$ keeps false during the duration $d$, and the accumulated execution time is less than the upper bound $T_o$. We first prove that for any $t \in [0, d]$, 
    $\|s_0[\xx \mapsto \mathbf{p}(t)]- s_0'[\xx \mapsto \xx + \lfloor \frac{t}{h}\rfloor \cdot h \PHI]\| < \varepsilon$. 
    The left hand side, denoted by $LHS$, satisfies ($U$ is the upper bound of $\ff(\xx)$):
    \[
    \begin{array}{ll}
         LHS&\leq \| s_0[\xx \mapsto \mathbf{p}(t)] - s_0[\xx \mapsto \mathbf{p}(h\cdot \lfloor \frac{t}{h}\rfloor)]\| + \\
         &\ \|s_0[\xx \mapsto \mathbf{p}(h\cdot \lfloor \frac{t}{h}\rfloor), s_0'[\xx \mapsto \xx + \lfloor \frac{t}{h}\rfloor\cdot h\PHI]\|  \\
         & \leq U \cdot h' +  M(h) < \frac{\varepsilon}{2} + \frac{\varepsilon}{2} = \varepsilon
    \end{array}\]
   Obviously, before time $\lceil \frac{d}{h} \rceil \cdot h$, denoted by $k$, the discretized states are in $N(B, \varepsilon)$, i.e. all states before $\lceil d \rceil \cdot h-h$ also satisfies $N^n(B, \varepsilon)$. In consequence,   
     process (\ref{eqn:ODE}) at least  has the following transitions:  
    \[((\ref{eqn:ODE}), s_0') (\xrightarrow{h} (\_, s_i') \rightarrow (\_, s_i'[\xx\mapsto\xx+h\PHI]))^{k-1} %(\xrightarrow{h'} (\_, s_k))
    \] 
    where $s_i' = s_{i-1}' [\xx\mapsto\xx+h\PHI]$ for any $i > 1$. If $N^n(B, \varepsilon)$ is still true for the $k$-th step, another transition proceeds and the final state is still within $\varepsilon$ of the precise continuous value. In the other direction, the first $k-1$ rounds are the same, but if $N^n(B, \varepsilon)$ is false for the $k$-th step, then at next step, $B$ must be false, (\ref{eqn:ODE}) terminates at previous one step, and the final values are still within $\varepsilon$. 
    
    \item If $(\langle\xx = \ff(\xx) \& B\rangle, s_0) \xrightarrow{\tau}(\pskip, s_0]$, when $s_0(B) = false$, $t_0\in (kh, (K+1)h]$. From the fact that the HCSP process is $(\delta, \epsilon)$-robustly safe, then there exists $\widetilde{t} \in (t_0,t_0+\delta)$ such that for all $s \in U (s_0[\xx \mapsto X(\widetilde{t}, s_0)], \epsilon)$, $U(s, \epsilon)$ makes $B$ false, and $h<\delta <2h$. It holds that $0 < \widetilde{t} - t_0 < 2h$.  Consider $\|\xx_{k+2} - X(\widetilde{t}, s_0)\|$, where $\xx_{k+2} = \xx + (k+2)h\PHI$, which is reduced to:
    %\[\|\xx_{k+2} - X((k+2)h, s_0)\| + \|X((k+2)h, s_0) - X(\widetilde{t}, s_0)\| < \varepsilon\]
      \[
    \begin{array}{l}
         \|\xx_{k+2} - X((k+2)h, s_0)\| + \\
       \quad   \|X((k+2)h, s_0) - X(\widetilde{t}, s_0)\|  
        < \varepsilon
    \end{array}\]
    Thus, $U(\xx_{k+2}, \epsilon) \in \neg B$, which means that $\xx_{k+2} \in N(\neg B, -\epsilon)$. According to process (\ref{eqn:ODE}), the condition for the $k+2$-th repetition violates, which if considered as the escaping point for the discretised version, more than one time step distance from the exact escaping point. After that, control variable $j$ is 0 and the rest repetition will terminate immediately. 
    \item After the upper time bound $T$, the discretised process is also after the $N$ repetition, both sides will do nothing.  
\end{itemize}

\item[2)]
We now prove the case for continuous interrupt, i.e. $\exempt{\evo{x}{e}{B}}{i\in I}{ch_i*}{p_i}$, denoted by $\textit{ODEI}$ below,  and the discrete process  (\ref{eqn:odeI}) are approximate bisimilar. For this case, the discretization of ODE is involved, but also it needs to consider the communication behavior. 

There are two cases, detailed as follows:
\begin{itemize}
    \item If $\textit{ODEI}$ performs a $d$ duration ($d > 0$), i.e. $B$ holds true  and no communication occurs within $d$ duration, or if it performs a $\tau$ transition because of the violation of $B$, then these two cases are very similar to  case continuous evolution. We can prove similarly that  (\ref{eqn:odeI}) can perform a corresponding sequence of transitions including a $h$ duration progress and the update of continuous states $x$ such that the final states are approximate bisimilar. 
    \item If $\textit{ODEI}$ performs a communication action, e.g. $\langle ch!, v\rangle$ (indexed by $k$), then there must exist a process in parallel such that it can perform the compatible  action  $\langle ch?, v\rangle$. There are two cases, the communication occurs at the time when $B$ is true, or when $B$ just turns false.  For the discrete process (\ref{eqn:odeI}),  $j_1$ is 1 for the first case  and $j_1$ is 0 for the second case. For both, $j_2$ is some $i$. So (\ref{eqn:odeI}) will execute to the interrupt to line 4 and then to line 6. The communication is enabled and the right continuation is chosen to execute. 
\end{itemize}

\end{itemize}

\subsection{Proof of discrete bisimulation}
\label{sec:proof-discrete-bisimulation}

We present the details of part of the bisimulation proof between discretised HCSP and the generated C code.  We first prove the primitives related to time delay and communications. 

\subsubsection{Delay Case}
We first consider the case of delay only. Given HCSP program $p$, we write the corresponding C program as $\overline{p}$. The bisimulation relation is given as follows. For each HCSP process $i$, one of the following two conditions hold:
\begin{enumerate}
\item[(1a)] The HCSP program is $p_i$ and the C program is $\overline{p_i}$, moreover $\mathit{localTime}(i)=\mathit{currentTime}$.
\item[(1b)] For some $d_i>0$, the HCSP program is $\mathsf{delay}(d_i); p_i$ and the C program is $\overline{p_i}$, moreover $\mathit{localTime}(i)   =\mathit{currentTime}+d_i$.
\end{enumerate}
Note the above condition implies $\mathit{currentTime}$ is less than or equal to $\mathit{localTime}(i)$ for any $i$. We add following invariant conditions on the C state:
\begin{enumerate}
\item[(2)] The global variable $\mathit{currentTime}$ equals the minimum of $\mathit{localTime}(i)$ for all $i$.
\item[(3)] For each thread $i$, its state is $\mathsf{WAITING}$ if $\mathit{localTime}(i)<\mathit{currentTime}$, otherwise its state is $\mathsf{RUNNING}$. The value of the condition variable corresponds to its state for each thread.
\end{enumerate}
Finally, the global time on the C side and HCSP side corresponds to each other:
\begin{enumerate}
\item[(4)] The value of $\mathit{currentTime}$ equals the global clock in HCSP.
\end{enumerate}

We now show that each execution of C program $\mathsf{delay}$ corresponds to either identity or the delay transition on the HCSP program. Note the entirety of $\mathsf{delay}$ is within a single critical region, so it can be considered atomic.

The operation performed by the C program $\mathsf{delay}(\mathit{tid},\mathit{secs})$ is as follows. It increments $\mathit{localTime}(\mathit{tid})$ by $\mathit{secs}$, then updates $\mathit{currentTime}$ to be the new minimum of $\mathit{localTime}(i)$. Any thread whose $\mathit{localTime}$ equals $\mathit{currentTime}$ afterwards is woken up.

There are two cases to consider, depending on whether $\mathsf{delay}(\mathit{tid},\mathit{secs})$ results in a change of $\mathit{currentTime}$.
\begin{itemize}
\item If $\mathit{currentTime}$ is not changed, then the transition on the C side corresponds to identity transition on the HCSP side. For thread $\mathit{tid}$, since the thread is allowed to run, by Condition 3), we have $\mathit{localTime}(i)=\mathit{currentTime}$ beforehand, so Condition 1a) holds, the HCSP program has form $\mathit{wait}(\mathit{secs});p$, and the C program has form $\mathit{delay}(\mathit{\mathit{tid},\mathit{secs}});\overline{p}$. After executing $\mathit{delay}(\mathit{tid},\mathit{secs})$, the C program becomes $\overline{p}$ and $\mathit{localTime}(i)$ becomes $\mathit{currentTime}+\mathit{secs}$, so Condition 1b) holds afterwards. The state of thread $\mathit{tid}$
is set to $\mathsf{WAITING}$, so Condition 3) holds afterwards. Finally Conditions 2) and 4) are unaffected.

\item If $\mathit{currentTime}$ is incremented by some $d$, then the transition on the C side corresponds to transition $\langle d,\emptyset\rangle$ on the HCSP side. The HCSP transition increments the global clock by $d$, and removes $\mathsf{delay}(d)$ from all processes. On the C side, for each process, since $\mathit{currentTime}$ is in incremented by $d$ and $\mathsf{delay}(d)$ is removed from HCSP program at the same time, Condition 1b) is either preserved or becomes Condition 1a). In the latter case, Condition 3) is preserved since thread states are updated and signals sent in function $\mathsf{updateCurrentTime}$. Condition 2) continues to hold due to the computation of $\mathit{currentTime}$ in function $\mathsf{updateCurrentTime}$.
\end{itemize}

\subsubsection{Communication Case}
Next, we consider the case of input and output statements. The C program for both input and output consist of two cases, depending on whether the other side is ready on entry to the function. Each case is divided into two atomic blocks. We first review the actions taken in each atomic block. Following the convention in the code listings, we let $\mathit{tid}$ be the current thread, $\mathit{i}$ be the other side of communication, and $\mathit{ch}$  be the channel.
\begin{itemize}
\item Input, other side ready: The first block sets the input thread of $ch$ to $\mathit{tid}$, sets the state of thread $i$ to $ch$, copies value from channel $ch$, and signals thread $i$. The second block updates the local time of $\mathit{tid}$ if necessary, updates the global clock, sets the thread state of $\mathit{tid}$ to $\mathsf{RUNNING}$, and finally sets the input thread of $ch$ to $-1$.

\item Input, other side not ready: The first block sets the input thread of $ch$ to $\mathit{tid}$, sets the state of thread $\mathit{tid}$ to $\mathsf{AVAILABLE}$,  assigns the local time to be the maximum \textsf{DBL\_MAX}, and update the current time. The second block updates the local time of $\mathit{tid}$, copies value from channel $ch$, sets the state of thread $\mathit{tid}$ to $\mathsf{RUNNING}$, sets the input thread of $ch$ to $-1$, and finally signals thread $i$.

\item Output, other side ready: The first block sets the output thread of $ch$ to $\mathit{tid}$, copies value sent to $ch$, sets the state of thread $i$ to $ch$, and signals thread $i$. The second block updates the local time of $\mathit{tid}$ if necessary, updates the global clock, sets the thread state of $\mathit{tid}$ to $\mathsf{RUNNING}$, and finally sets the input thread of $ch$ to $-1$.

\item Output, other side not ready: The first block sets the output thread of $ch$ to $\mathit{tid}$, copies value sent to $ch$, sets the state of thread $\mathit{tid}$ to $\mathsf{AVAILABLE}$, assigns the local time to be the maximum \textsf{DBL\_MAX}, and update the current time. The second block updates the local time of $\mathit{tid}$,   sets the state of thread $\mathit{tid}$ to $\mathsf{RUNNING}$,  sets the output thread of $ch$ to $-1$, and finally signals thread $i$.
\end{itemize}

From these descriptions, we can conclude the following pattern: If the other side is ready, the first block performs the communication and signals to the other side of this fact, the second block performs other cleanup. If the other side is not ready, the first block sets appropriate variables and then wait for the other side to become ready, the second block updates local time and performs other cleanup.

We name the second block of $\mathsf{input}$ when the other side is ready (resp. other side is not ready) to be $\mathsf{input_a}$ (resp. $\mathsf{input_b}$). Likewise, we name the second block of $\mathsf{output}$ when the other side is ready (resp. other side is not ready) to be $\mathsf{output_a}$ (resp. $\mathsf{output_b}$). These notations will be used in the examples as well as when defining the bisimulation relation below.

Intuitively, each communication executes in one of the following two patterns:
\begin{itemize}
\item The input side is ready first, and enters into waiting for input mode. Then the output side provides the value, the input side receives it and signals to the output side on the success of communication. This corresponds to the following four steps:
\[
\begin{array}{l}
\mathsf{input},\mathsf{output} \longrightarrow
\mathsf{input_b},\mathsf{output} \longrightarrow \\
\mathsf{input_b},\mathsf{output_a} \longrightarrow
\mathsf{skip},\mathsf{output_a} \longrightarrow
\mathsf{skip},\mathsf{skip}
\end{array}
\]

\item The output side is ready first, and enters into waiting for output mode. Then the input side receives the value and signals to the output side, finally the output side signals back to the input side on the success of communication. This corresponds to the following four steps:
\[
\begin{array}{l}
\mathsf{input},\mathsf{output} \longrightarrow
\mathsf{input},\mathsf{output_b} \longrightarrow \\
\mathsf{input_a},\mathsf{output_b} \longrightarrow
\mathsf{input_a},\mathsf{skip} \longrightarrow
\mathsf{skip},\mathsf{skip}
\end{array}
\]
\end{itemize}

Note the four steps may be interrupted by other actions in between. An example of the first case is given in Example~\ref{ex:communication}, where a waiting action interrupts between the first and second step.

The key idea of the proof is to associate the step where the input side receives the transmitted value to the transition on the HCSP side. In the case when the input side is ready first, this is the third of the four transitions. In the case when the output side ready first, this is the second of the four transitions.

Hence, the bisimulation relation for the case when input side is ready first is given as follows. For each HCSP process $i$, one of the following conditions hold.
\begin{enumerate}
\item[(1a)] The HCSP program is $p_i$ and the C program is $\overline{p_i}$, moreover the state is $\mathsf{RUNNING}$,  and $\mathit{localTime}(i)=\mathsf{currentTime}$.

\item[(1b)] The HCSP program is $ch?x;p_i$ and the C program is $\mathsf{input_b};\overline{p_i}$, moreover the state is $\mathsf{AVAILABLE}$,   and $\mathit{localTime}(i)$ is $\mathsf{DBL\_MAX}$. Finally the input side of channel $ch$ is set to the current thread.

\item[(1c)] The HCSP program is $ch!e;p_i$ and the C program is $\mathsf{output_a};\overline{p_i}$, moreover the thread is blocked, the output side of channel $ch$ is set to the current thread, and the value of $ch$ is set to the value of expression $e$ at the current state.

\item[(1d)] The HCSP program is $p_i$ and the C program is $\mathsf{output_a};\overline{p_i}$, moreover the thread is unblocked, the value of $x$ is set to the value of $ch$. 
\end{enumerate}

The idea of the proof is to show the following:
\begin{enumerate}
\item[(1)] The atomic transition from $\mathsf{input}$ to $\mathsf{input_b}$ carries a local state satisfying Condition 1a) to one satisfying Condition 1b), without any changes on the HCSP side.
\item[(2)] The atomic transition from $\mathsf{input_b}$ to $\mathsf{skip}$ carries a local state satisfying Condition 1b) to one satisfying Condition 1a), while the communication is carried out on the HCSP side. Hence the variable $x$ is assigned, and the output thread changes from satisfying Condition 1c) to satisfying Condition 1d).
\item[(3)] The atomic transition from $\mathsf{output}$ to $\mathsf{output_a}$ carries a local state satisfying Condition 1a) to one satisfying Condition 1c), without any changes on the HCSP side.
\item[(4)] The atomic transition from $\mathsf{output_a}$ to $\mathsf{skip}$ carries a local state satisfying Condition 1d) to one satisfying Condition 1a), without any changes on the HCSP side.
\end{enumerate}

\subsubsection{General case $\mathsf{wait\_comm}$ for interrupt}
\label{sec:interruptproof}

We now consider the following HCSP primitive, added for the discretization of interrupt:
\[ i := \mathsf{wait\_comm}(d,\mathit{chs}).\]
This primitive waits for time $d$ as well as communications $\mathit{chs}$. If one of the communications in $\mathit{chs}$ is ready before time $d$, then the communication is carried out, and $i$ is assigned its index in $\mathit{chs}$. Otherwise, $i$ is assigned $-1$.

Note $\mathsf{delay}$, $\mathsf{input}$ and $\mathsf{output}$ are all special cases of $\mathsf{wait\_comm}$. The $\mathsf{delay}(d)$ operation corresponds to $\mathsf{wait\_comm}(d,[])$. The input operation $ch?x$ corresponds to $\mathsf{wait\_comm}(\infty,[ch?x])$, and the output operation $ch!e$ corresponds to $\mathsf{wait\_comm}(\infty,[ch!e])$.

The translation of the above primitive to C code have the following structure.
\begin{itemize}
\item First, for each communication in $\mathit{chs}$, update the corresponding $\mathsf{channelInput}$ and $\mathsf{channelOutput}$ arrays. In the case of output, update the channel value as well. Then collect the list of matching communications that are ready.

\item If at least one matching communication is ready, nondeterministically choose one of them and proceed to carry out the communication as in the case of $\mathit{input}$ or $\mathit{output}$. At the end, reset the information in $\mathsf{channelInput}$ and $\mathsf{channelOutput}$.

\item If no matching communication is ready, the local time of the thread is incremented by $\mathit{secs}$ and the current time is updated. If the local time is now greater than current time, the thread enters $\mathsf{WAITING\_AVAILABLE}$ mode and blocks.

\item After the thread is unblocked, if the thread is in $\mathsf{AVAILABLE}$ mode, this means the time limit is reached first, then the function returns $-1$.

\item After the thread is unblocked, if the thread state contains a channel, this means some communication happened first, then the corresponding operations for the second part of the communication is performed. In particular, this updates local time to the time of communication, which is the local time of the other thread.
\end{itemize}

We identify the following atomic action blocks for $\mathsf{wait\_comm}$. The block $\mathsf{wait\_comm_a}_i$ is the second atomic block when the $i$'th communication in $\mathit{chs}$ is initially ready and chosen. The block $\mathsf{wait\_comm_b}_i$ is the second atomic block when the $i$'th communication happened during waiting. Finally, the entire block $\mathsf{wait\_comm}$ may be executed in the case when the time limit $\mathit{secs}$ is reached without communication. The possible execution orders for $\mathsf{wait\_comm}$ are the following:
\begin{enumerate}
\item[(1)] $\mathsf{wait\_comm} \rightarrow \mathsf{wait\_comm_a}_i \rightarrow \mathsf{skip}$ for some $0\le i<|\mathit{chs}|$, the return value is $i$.
\item[(2)] $\mathsf{wait\_comm} \rightarrow \mathsf{wait\_comm_b}_i \rightarrow \mathsf{skip}$ for some $0\le i<|\mathit{chs}|$, the return value is $i$.
\item[(3)] $\mathsf{wait\_comm} \rightarrow \mathsf{skip}$, and the return value is $-1$.
\end{enumerate}
The first and third cases are very similar to the communication case and the delay case proved previously, so we will not list the proofs for them here. Below we give the proof of the second case, for which the other sides of all channels are not ready at the beginning and then one of them  gets ready after waiting for some time less than the time limit $h$. This corresponds to the following steps (assume the corresponding channel is $ch$ and thus we omit the channel subscript for presentation):
\[
\begin{array}{l}
\mathsf{wait\_comm},(\mathsf{delay; output}) \longrightarrow\\
\mathsf{\mathsf{wait\_comm}_b},\mathsf{(delay; output)} \longrightarrow \\
\mathsf{\mathsf{wait\_comm}_b},\mathsf{output} \longrightarrow
\mathsf{\mathsf{wait\_comm}_b},\mathsf{output_a} \longrightarrow \\
\mathsf{skip},\mathsf{output_a} \longrightarrow
\mathsf{skip},\mathsf{skip}
\end{array}
\]

The bisimulation relation for this relation is given as follows. For each HCSP process $i$, one of the following conditions hold:
\begin{enumerate}
\item[(1a)] The HCSP program is $p_i$ and the C program is $\overline{p_i}$, moreover the state is $\mathsf{RUNNING}$,  and $\mathit{localTime}(i)=\mathsf{currentTime}$.

\item[(1b)] The HCSP program is $\mathit{wait\_comm};p_i$ and the C program is $\mathsf{wait\_comm_b};\overline{p_i}$, moreover the state is $\mathsf{WAITING\_AVAILABLE}$,   and $\mathit{localTime}(i)$ is added by $\mathsf{h}$. Finally the output side of channel $ch$ is set to the current thread.

\item[(1c)] The HCSP program is $ch!e; p_i$ and the C program is $\mathsf{output_a};\overline{p_i}$, moreover the thread is blocked, the output side of channel $ch$ is set to the current thread, and the value of $ch$ is set to the value of expression $e$ at the current state.

\item[(1d)] The HCSP program is $p_i$ and the C program is $\mathsf{output_a};\overline{p_i}$, moreover the thread is unblocked, the value of $x$ is set to the value of $ch$. 
\end{enumerate}

The idea of the proof is to show the following:
\begin{enumerate}
\item[(1)] The atomic transition from $\mathsf{wait\_comm}$ to $\mathsf{\mathsf{wait\_comm}_b}$ carries a local state satisfying Condition 1a) to one satisfying Condition 1b), without any changes on the HCSP side.
\item[(2)] The atomic transition from $\mathsf{delay; output}$ to $\mathsf{output}$ carries a local state satisfying Condition 1a) to one still satisfying Condition 1a), corresponding to a wait statement on the HCSP side.
\item[(3)] The atomic transition from $\mathsf{output}$ to $\mathsf{output_a}$ carries a local state satisfying Condition 1a) to one satisfying Condition 1c), without any changes on the HCSP side.
\item[(4)] The atomic transition from $\mathsf{\mathsf{wait\_comm}_b}$ to $\mathsf{skip}$ carries a local state satisfying Condition 1b) to one satisfying Condition 1a), while the communication is carried out on the HCSP side. Hence the variable $x$ is assigned, and the output thread changes from satisfying Condition 1c) to satisfying Condition 1d).
\item[(5)] The atomic transition from $\mathsf{output_a}$ to $\mathsf{skip}$ carries a local state satisfying Condition 1d) to one satisfying Condition 1a), without any changes on the HCSP side.
\end{enumerate}

\subsubsection{Other compound cases}
Above we have proved the primitives related to time delay and communication. Other cases such as in (\ref{eqn:odeI}), the expressions, and the statements such as assignments, conditional, repetition with determined number of times, sequential composition, and so on, are transformed to C directly without any essential changes and the bisimulation relation between them and the generated C code can be proved trivially based on structural induction.

\subsection{Examples to illustrate the proof}
\begin{enumerate}
\item[(1)] Figure~\ref{ex:delay} shows an example with delay only. The initial HCSP program is
\[ \mathsf{wait}(10) ~\|~ \mathsf{wait}(20) ~\|~ \mathsf{wait}(30) \]

\item[(2)]Figure~\ref{ex:communication} shows an example with input, output and delay. The initial HCSP program is
\[ ch_1?x ~\|~ \mathsf{wait}(10); ch_1!3 \]
\end{enumerate}

\begin{figure*}
\centering
\begin{tabular}{c c c c|c}
$p_1$ & $p_2$ & $p_3$ & $\mathit{cTime}$ & $\mathsf{HCSP}$
\\ \hline
$\langle \mathsf{delay}(10), 0, R \rangle$ &
$\langle \mathsf{delay}(20), 0, R \rangle$ &
$\langle \mathsf{delay}(30), 0, R \rangle$ &
$0$ &
$\mathsf{wait}(10) ~\|~ \mathsf{wait}(20) ~\|~ \mathsf{wait}(30)$
\\ \hline
$\langle\mathsf{skip},10,W\rangle$ &
$\langle\mathsf{delay}(20),0,R\rangle$ &
$\langle\mathsf{delay}(30),0,R\rangle$ &
$0$ &
$\mathsf{wait}(10) ~\|~ \mathsf{wait}(20) ~\|~ \mathsf{wait}(30)$
\\ \hline
$\langle\mathsf{skip},10,W\rangle$ &
$\langle\mathsf{skip},20,W\rangle$ &
$\langle\mathsf{delay}(30),0,R\rangle$ &
$0$ &
$\mathsf{wait}(10) ~\|~ \mathsf{wait}(20) ~\|~ \mathsf{wait}(30)$
\\ \hline
$\langle\mathsf{skip},10,R\rangle$ &
$\langle\mathsf{skip},20,W\rangle$ &
$\langle\mathsf{skip},30,W\rangle$ &
$10$ &
$\mathsf{skip} ~\|~ \mathsf{wait}(10) ~\|~ \mathsf{wait}(20)$
\\ \hline
$\langle\mathsf{skip},10,S\rangle$ &
$\langle\mathsf{skip},20,R\rangle$ &
$\langle\mathsf{skip},30,W\rangle$ &
$20$ &
$\mathsf{end} ~\|~ \mathsf{skip} ~\|~ \mathsf{wait}(10)$
\\ \hline
$\langle\mathsf{skip},10,S\rangle$ &
$\langle\mathsf{skip},20,S\rangle$ &
$\langle\mathsf{skip},30,R\rangle$ &
$30$ &
$\mathsf{end} ~\|~ \mathsf{end} ~\|~ \mathsf{skip}$
\\ \hline
$\langle\mathsf{skip},10,S\rangle$ &
$\langle\mathsf{skip},20,S\rangle$ &
$\langle\mathsf{skip},30,S\rangle$ &
$30$ &
$\mathsf{end} ~\|~ \mathsf{end} ~\|~ \mathsf{end}$
\end{tabular}
\caption{Example of bisimulation between the execution of C program and HCSP program, for the starting process $\mathsf{wait}(10) ~\|~ \mathsf{wait}(20) ~\|~ \mathsf{wait}(30)$. Each C thread is specified by the triple $\langle p,lt,st\rangle$, where $p$ is the program that remains to be executed, $lt$ is the local time, and $st$ is the state.}
\label{ex:delay}
\end{figure*}

\begin{figure*}
\centering
\begin{tabular}{c c c c|c}
$p_1$ & $p_2$ & $ch_1$ & $\mathit{cTime}$ & $\mathsf{HCSP}$
\\ \hline
$\langle \mathsf{input}(ch_1), 0, R, 0\rangle$ &
$\langle \mathsf{delay}(10);\mathsf{output}(ch_1), 0, R, 0\rangle$ &
$\langle -1, -1, \mathsf{null}\rangle$ &
0 &
$ch_1?x ~\|~ \mathsf{wait}(10); ch_1!3$
\\ \hline
$\langle \mathsf{input_b}(ch_1), 0, A, 1\rangle$ &
$\langle \mathsf{delay}(10);\mathsf{output}(ch_1), 0, R, 0\rangle$ &
$\langle p_1, -1, \mathsf{null}\rangle$ &
0 &
$ch_1?x ~\|~ \mathsf{wait}(10); ch_1!3$
\\ \hline
$\langle \mathsf{input_b}(ch_1), 0, A, 1\rangle$ &
$\langle \mathsf{output}(ch_1), 10, R, 0\rangle$ &
$\langle p_1, -1, \mathsf{null}\rangle$ &
10 &
$ch_1?x ~\|~ ch_1!3$
\\ \hline
$\langle \mathsf{input_b}(ch_1), 0, ch_1, 1\rangle$ &
$\langle \mathsf{output_a}(ch_1), 10, R, 0\rangle$ &
$\langle p_1, p_2, 3\rangle$ &
10 &
$ch_1?x ~\|~ ch_1!3$
\\ \hline
$\langle \mathsf{skip}, 10, R, 0\rangle$ &
$\langle \mathsf{output_a}(ch_1), 10, R, 0\rangle$ &
$\langle -1, p_2, 3\rangle$ &
10 &
$\mathsf{skip} ~\|~ \mathsf{skip}$
\\ \hline
$\langle \mathsf{skip}, 10, R, 0\rangle$ &
$\langle \mathsf{skip}, 10, R, 0\rangle$ &
$\langle -1, -1, \mathsf{null}\rangle$ &
10 &
$\mathsf{skip} ~\|~ \mathsf{skip}$
\end{tabular}
\caption{Example of bisimulation between the execution of C program and HCSP program, for the starting process $ch_1?x ~\|~ \mathsf{wait}(10); ch_1!3$. Each C thread is specified by the quadruple $\langle p,lt,st,pw\rangle$, where $p$ is the program that remains to be executed, $lt$ is the local time, $st$ is the state, and $pw$ is the boolean variable $\mathsf{permWait}$. Each channel is specified by the triple $\langle in,out,val\rangle$, where $in$ is the input side, $out$ is the output side, and $val$ is the value to be transmitted.}
\label{ex:communication}
\end{figure*}

\section{Case Study}\label{CaseStudy}

We adopt the realistically-scaled Automatic Cruise Control System (ACCS for short) from~\cite{DBLP:journals/tcs/XuWZJTZ22} as the case study. \newcomment{The HCSP model of ACCS is moderately large and covers all the concerned features of HCSP 
and it is robustly safe (Sect.~\ref{sec:robustly-safe}) and hence can be discretized using our approach.}
In this section, we translate this HCSP model to the C code and compare its execution with the simulation of the HCSP model~\cite{DBLP:journals/tcs/XuWZJTZ22} and the execution of the C code generated directly from the original $\textsc{AADL}\oplus\textsc{S/S}$ model~\cite{DBLP:conf/utp/ZhanLWTXZ19}.

\oomit{
\begin{figure*}
\centering
\includegraphics[scale=0.45]{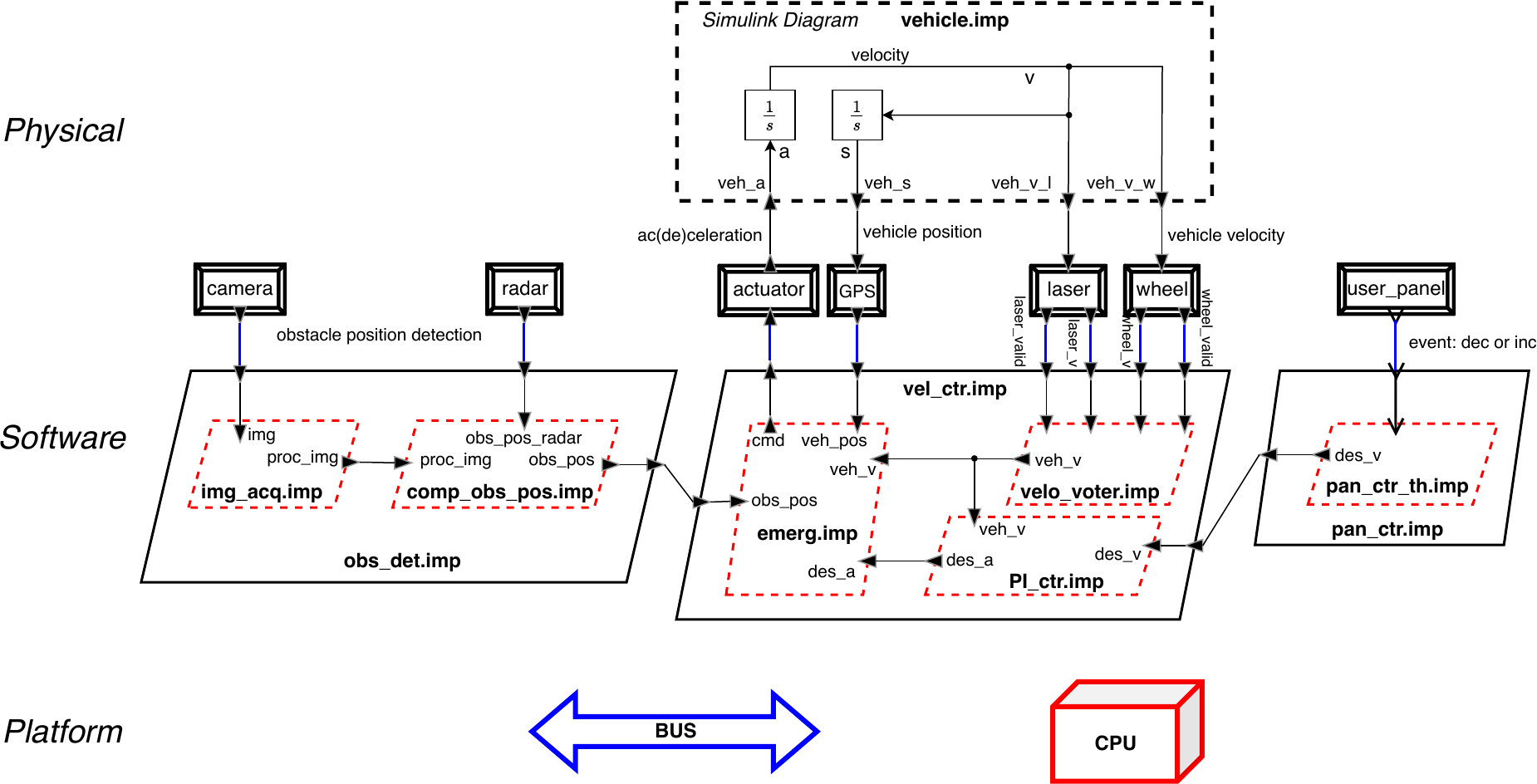}
\caption{Automatic Cruise Control System (borrowed from~\cite{DBLP:journals/tcs/XuWZJTZ22})}
\label{fig:CCS}
\end{figure*}
}

\subsection{The Automatic Cruise Control System}\label{ACCS_arch}

The architecture of ACCS consists of three layers. The physical layer is the physical vehicle. The software level defines control of the system and it contains three processes for obstacle detection, velocity control, and panel control, and each process is composed of several threads. These processes interact with the environment (the physical layer) through devices. The platform layer consists of a bus and a processor. The connections between processes and devices could be bound to the bus and all the threads are bound to the processor, with HPF (High-Priority-First) scheduling policy.

% The execution of ACCS is as follows. A vehicle is placed at the starting point initially and the driver can accelerate (\texttt{inc}) and decelerate (\texttt{dec}) the vehicle by the \texttt{user\_panel}. Thread \texttt{pan\_ctr\_th} deals with the commands from the driver and then sends desired velocities (\texttt{des\_v}) to the discrete PI controller (thread \texttt{PI\_ctr}). Meanwhile, process \texttt{obs\_det} detects the obstacles ahead by a \texttt{camera} and a \texttt{radar} and provides the velocity controller process (\texttt{vel\_ctr}) with the real-time position of obstacle (\texttt{obs\_pos}). Thread \texttt{velo\_voter} in process \texttt{vel\_ctr} monitors the velocity of the vehicle using \texttt{laser} and another device located on one \texttt{wheel} of the vehicle and produces the real-time velocity of the vehicle (\texttt{veh\_v}) to the discrete PI controller \texttt{PI\_ctr} and the emergency control thread (\texttt{emerg}). Based on the real-time velocity of vehicle and the desired velocity received, \texttt{PI\_ctr} computes a desired acceleration (\texttt{des\_a}) which will be sent to \texttt{emerg}. Finally, \texttt{emerg} collects the real-time position (\texttt{veh\_pos} by \texttt{GPS}) and velocity of the vehicle, the desired velocity set by the driver, and the real-time position of the obstacle to work out a \texttt{command}, by some emergency control strategy, which will update the acceleration of the vehicle through \texttt{actuator}. The vehicle moves according to the new acceleration and the above procedure repeats. 

\subsection{Translation from HCSP to C}

In work~\cite{DBLP:conf/utp/ZhanLWTXZ19}, combined models of $\textsf{AADL}\oplus\textsf{S/S}$ are translated to C code directly. \oomit{The translation is done by the following steps: (1) the AADL part is translated to C following the execution semantics of AADL; (2) the Simulink/Stateflow part is translated using the existing C code generation facility in Matlab; (3) the architecture part is implemented by a library in C that includes thread scheduling protocols and so on; and (4) the combined model is translated by integrating the C codes generated from the above three parts. However, the interaction (communication) between components is implemented by shared variables (without \textsf{pthreads}) and the correctness of the translation has not been proved.
}
In this paper, 
%we generate the C code of ACCS from its HCSP model obtained from~\cite{DBLP:journals/tcs/XuWZJTZ22}. The HCSP model is specified by a formal language and therefore verifiable. Thus, the C code generated from HCSP models is more reliable than~\cite{DBLP:conf/utp/ZhanLWTXZ19}.
we generate the C code of the case study from its HCSP model which has been verified in~\cite{DBLP:journals/tcs/XuWZJTZ22}. Since the correctness of the translation is guaranteed (Sect.~\ref{sec:correct}), the generated code, especially the code for the controller (the velocity control mentioned in Sect.~\ref{ACCS_arch}), satisfies the safety requirement and therefore is reliable.
We use $\mathsf{pthreads}$ to implement the communication between processes and the correctness of the code generation has been proved in Sect.~\ref{sec:correct}. The generated C code is of 3500--4000 lines, longer than the C code (about 2500 lines) generated by \cite{DBLP:conf/utp/ZhanLWTXZ19}. 
%In spite of its reliability and correctness, the generated code is, however, verbose: 3500--4000 lines of C code (while about 2500 lines by the translation of work \cite{DBLP:conf/utp/ZhanLWTXZ19}). 
One major reason is that the generated C code in this paper is approximately bisimilar to the original $\textsf{AADL}\oplus\textsf{S/S}$ model of ACCS, which means the detailed behaviours of ACCS are reflected in the C code and vice versa, while it is not the case for~\cite{DBLP:conf/utp/ZhanLWTXZ19} where no such bisimulation can be guaranteed. 
%In addition, the generated C code of this paper is more readable than~\cite{DBLP:conf/utp/ZhanLWTXZ19}, because in the latter, the C code of the Simulink part of ACCS is generated by the tool in Matlab, which introduces 

\subsection{Comparison}

We consider the following scenario. At the beginning, the driver pushes the \texttt{inc} button three times with time interval $0.5$s in between to set a desired speed to $3$m/s. After $30$s, the driver pushes the \texttt{dec} button twice in $0.5$s time intervals to decrease the desired speed. On the obstacle side, we assume that the obstacle appears at time $10$s and position $35$m, then moves ahead with velocity $2$m/s, before finally moving away at time $20$s and position $55$m. 

\oomit{
We adopt the parameters setting of Table 1 in~\cite{DBLP:journals/tcs/XuWZJTZ22} for the threads and devices in the ACCS of Fig.~\ref{fig:CCS}, except that the deadline of thread \texttt{pan\_ctr\_th} is shrunk to $50$ms in this paper.
We compare our results with the work~\cite{DBLP:conf/utp/ZhanLWTXZ19} and the simulation of the HCSP model of ACCS~\cite{DBLP:journals/tcs/XuWZJTZ22}. It should be noted that bus latency is not considered in~\cite{DBLP:conf/utp/ZhanLWTXZ19}. Thus, to make it fair, we only consider the case with on bus latency, i.e., we ignore the \texttt{bus} component in Fig.~\ref{fig:CCS}. Actually, \texttt{bus} has no impact on the study, as we only concern the translation from HCSP to C in this paper and each \texttt{bus} in the original $\textsf{AADL}\oplus\textsf{S/S}$ model will be translated to a specific process in the obtained HCSP model by~\cite{DBLP:journals/tcs/XuWZJTZ22}.
}

\begin{figure}[h]
\begin{minipage}{0.49\linewidth}
\centering
\includegraphics[width=\linewidth]{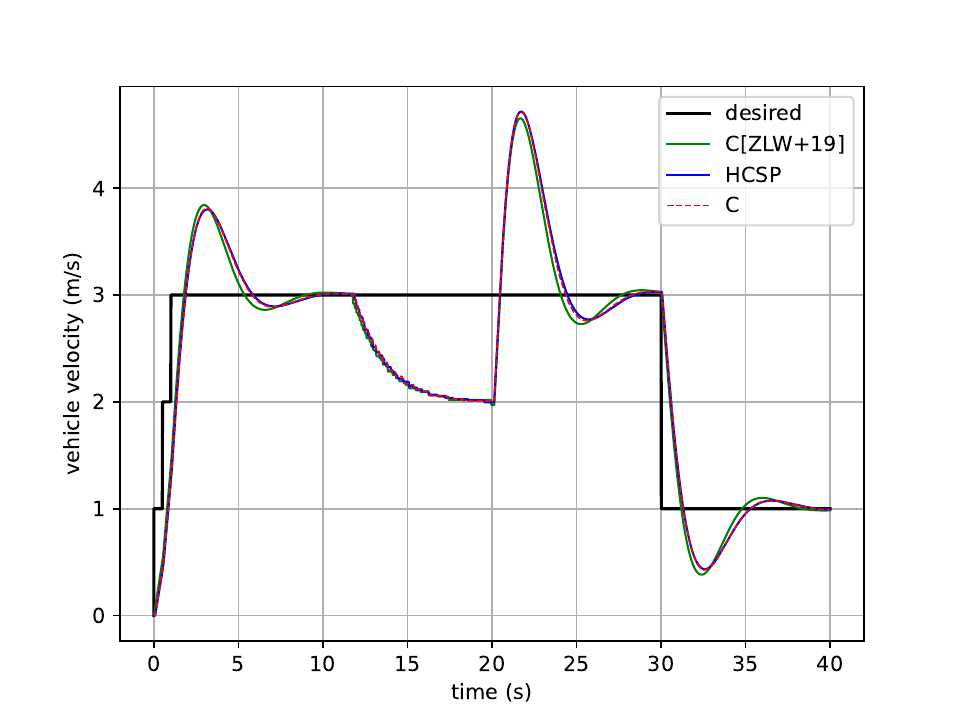}
\end{minipage}
\begin{minipage}{0.49\linewidth}
\centering
\includegraphics[width=\linewidth]{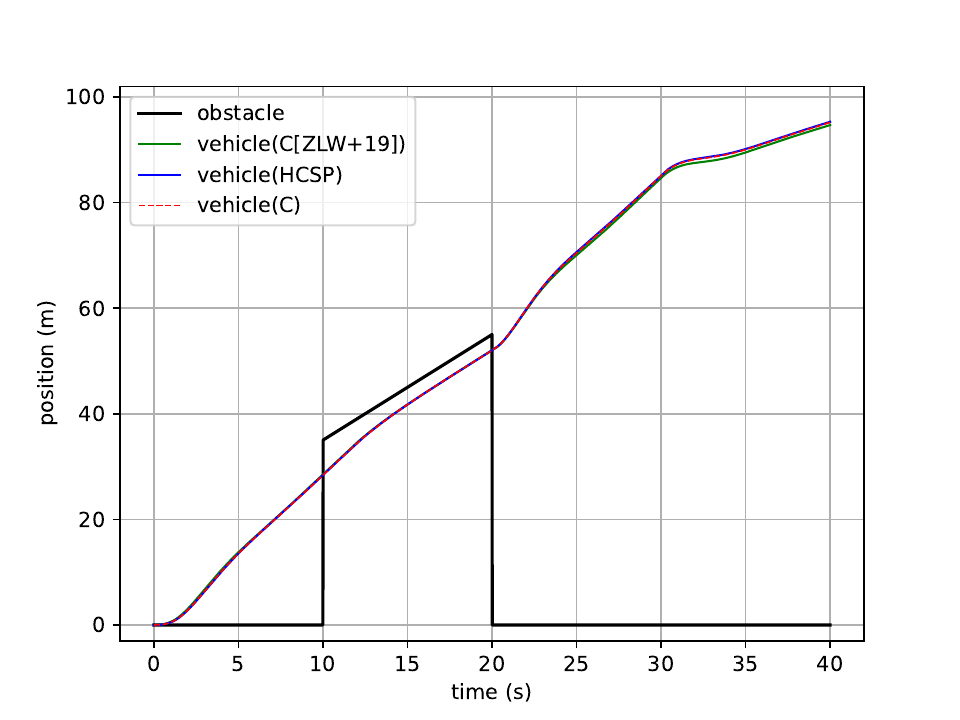}
\end{minipage}
\caption{Comparison of execution results}
\label{fig:veh_v_s}
\end{figure}

The top of Fig.~\ref{fig:veh_v_s} shows the execution results of the vehicle speed, where the black line denotes the desired velocity set by the driver, and the red, green, and blue lines denote the results of our work, the work of~\cite{DBLP:conf/utp/ZhanLWTXZ19}, and the simulation of the HCSP model of ACCS~\cite{DBLP:journals/tcs/XuWZJTZ22}, respectively. We can see that the execution result (red line) of the generated C code is almost the same with the simulation (blue line) of its HCSP model.
\newcomment{Specifically, the average relative error (ARE) between the time series of the velocity generated from the HCSP model and its C code is $0.138\%$ with the variance $4.686\times10^{-5}$.}
Besides, we can also observe that there is negligible difference between the results  of the C code generated by~\cite{DBLP:conf/utp/ZhanLWTXZ19} (green line) and the C code generated in this paper (red line):
\newcomment{the ARE is $0.182\%$ with the variance $4.232\times10^{-3}$.}
Readers can refer to~\cite{DBLP:journals/tcs/XuWZJTZ22} for more details about this example.

\newcomment{It should be noted that although the generated C code in this paper is longer and somewhat less efficient than the C code generated directly from the original graphical model~\cite{DBLP:conf/utp/ZhanLWTXZ19}, the former is more reliable because it is translated from the (formal) HCSP model, which can be verified by tools like HHLPy~\cite{DBLP:conf/fm/ShengBZ23}, and moreover, the correctness of the translation can be guaranteed (Sect.~\ref{sec:correct}).}
%These results are quite similar, though some negligible latency, largely due to our discretization of the ODEs, can be observed in the result of our work (red line). 

\oomit{
From all these results, we can see that the vehicle accelerates to the desired speed ($3$m/s) in $10$s. The fluctuation during $[2\text{s},10\text{s}]$ reflects feature of PI controllers.
%: the vehicle under control rushes over the desired velocity and then returns about desired velocity soon. 
It then decelerates to avoid the collision onto the obstacle ahead. After the obstacle moves away (at $20$s), the vehicle accelerates again to the desired speed. At $30$s, the driver pushes the \texttt{dec} button to adjust the desired velocity to $1$m/s and we can see that the vehicle decelerates to $1$m/s in about $6$s under the PI controller. The positions of the vehicle and of the obstacle with respect to time are shown at the bottom of Fig.~\ref{fig:veh_v_s}.
%\vspace{-3mm}
}

\section{Conclusions}\label{Conclusions}

This paper presents a formally verified C code generator from hybrid systems modelled using HCSP. In the transformation from HCSP to C, two main issues are addressed: the discretisation of continuous evolution and interrupts, and the realisation of synchronized communication of HCSP in concurrent C with the use of mutexes. For the first problem, we provide a method to compute a discretised time step such that under some robustness conditions, the continuous HCSP processes and the corresponding discrete ones are approximate bisimilar within the given precision. For the second problem, we propose a new notion of bisimulation relation: each transition step of HCSP corresponds to a sequence of atomic transitions of C, among which the substantial step to perform the equivalent transition in HCSP is determined. By transitivity of bisimulation relations, the correctness of the transformation from HCSP to C is guaranteed. Finally, we investigate a case study on Automatic Cruise Control System and its code generation using our approach.
%and furthermore show the correctness of its execution results through simulation. 
For future work, we consider to apply our approach to more practical case studies in industry.

\bibliographystyle{alpha}
\bibliography{sample}

\clearpage
\section*{Appendix}

\subsection{Part of The Generated C code for HCSP}

\subsubsection{Delay}
The time delay statement, $\pwait\ sec$, is implemented by the following C function \texttt{delay(tid, secs)}, where \texttt{tid} is the ID of current thread and \texttt{secs} the delayed time duration.
\begin{lstlisting}[style=mystyle,language=custom]
void delay(int tid, double secs) {
    lock mutex;
    localTime[tid] += secs;
    updateCurrentTime(tid);
    if (localTime[tid] > currentTime) {
        threadState[tid] = WAITING;
        cwait cond[tid] mutex;
    }
    unlock mutex;
}
\end{lstlisting}
It first locks \texttt{mutex}, and then increases the local clock of \texttt{tid} by \texttt{secs}. As the local clock increases, the global clock (which is the minimum of all local clocks) needs to be updated, and meanwhile, releases the threads on waiting states to execute, defined by  \texttt{updateCurrentTime(tid)}. After this, the state of thread \texttt{tid} is set to be \texttt{WAITING}, then waits till other threads catch up with it.

For \texttt{updateCurrentTime(tid)}, lines (2-8) assigns \texttt{minLocal} to be the minimum of the local clocks of active threads that are not waiting for communication. 
If the global clock is less than \texttt{minLocal}, it is progressed to be equal to 
\texttt{minLocal} (Line 9). Lines (12-21) checks all threads (other than current thread): if the local clock of the thread is equal to the global clock and meanwhile its state is \texttt{WAITING}, then update it to be \texttt{RUNNING} and wake it up by sending a signal; and if the state of the thread is \texttt{WAITING\_AVAILABLE}, then update it to be \texttt{AVAILABLE}, meaning that the thread still needs to wait for some communication. 

{\small
\begin{lstlisting}[style=mystyle,language=custom]
void updateCurrentTime(int tid) {
    double minLocal = DBL_MAX;
    for (int i = 0; i < numThread; i++) {
        if (threadState[i] != STOPPED &&
            localTime[i] < minLocal) {
            minLocal = localTime[i];
        }
    }
    if (currentTime < minLocal) {
        currentTime = minLocal;
    }
    for (int i = 0; i < numThread; i++) {
        if (i != tid && localTime[i] == currentTime &&
            threadState[i] == WAITING) {
            threadState[i] = RUNNING;
            signal cond[i];
        }
        if (i != tid && localTime[i] == currentTime &&
            threadState[i] ==  WAITING_AVAILABLE) {
            threadState[i] =  AVAILABLE;
            signal cond[i];
        }
    }
}
\end{lstlisting}
}

\subsubsection{Input}
The input $ch?x$ is transformed into $\texttt{input (tid, chi)}$, where $\texttt{tid}$ is the ID of current thread and $\texttt{chi}$ is the input channel corresponding to $ch?x$.  

%\comment{Why here, wait\_available to running when receiving available signal? But in  updateCurrentime,  wait\_available to available when receiving wait signal?}

\begin{lstlisting}[style=mystyle,language=custom]
void input(int tid, Channel chi) {
  lock mutex;
  channelInput[chi.channelNo] = tid;
  int i = channelOutput[chi.channelNo];
  if (i != -1 & (threadState[i] == AVAILABLE ||
      threadState[i] == WAITING_AVAILABLE)) { 
    threadState[i] = chi.channelNo;
    copyFromChannel(chi);
    signal cond[i];
    cwait cond[tid] mutex;
    channelInput[chi.channelNo] = -1;
  }
  else {
    threadState[tid] = AVAILABLE;
    localTime[tid] = DBL_MAX;
    updateCurrentTime(tid);
    cwait cond[tid] mutex;
    copyFromChannel(chi);
    threadState[tid] = RUNNING;
    localTime[tid] = localTime[i];
    channelInput[chi.channelNo] = -1;
    signal cond[i];
  }
  unlock mutex;
}
\end{lstlisting}
At the beginning (lines 2-4), thread \texttt{tid} first locks \texttt{mutex},
sets current  thread \texttt{tid} to be ready for the input side of channel \texttt{chi.channelNo} (i.e. the ID shared by \texttt{chi} and the compatible output channel, denoted by $cid$ below); Let  \texttt{i} be the  thread of the compatible output, recored in \texttt{channelOutput}. If the sender is already ready on thread \texttt{i} (indicated by \texttt{i != -1}) and furthermore \texttt{i} is on a state waiting for communication, then the following sequence of actions is performed (lines 5 - 13): the state of sender thread \texttt{i} is set to be the channel ID \texttt{chi} (line 7), the input channel \texttt{chi} copies value from the channel content that is already written by the output side and assigns it to variable $x$ (line 8); then, thread \texttt{tid} issues signal \texttt{cond[i]} to the sender \texttt{i} (line 9), releases \texttt{mutex} and waits on the reply signal from the sender (line 10). After receiving the signal from the sender, the whole communication  completes, resulted in setting the thread of input \texttt{chi} to be -1 (indicating it becomes unready). 

Otherwise, if the sender is not ready to execute (lines 13-224): first sets the state of thread \texttt{tid} to be \texttt{Available} to indicate that it is waiting for the compatible output (line 15), make \texttt{localtime} to be infinity (line 16); after that, releases lock \texttt{mutex} and waits for the sender (line 18); as soon as it is woken up again, meaning that the sender gets ready,
  the value can be obtained from the channel content array (line 19); the local time is updated, the state of \texttt{tid} is reset to \texttt{RUNNING}, and \texttt{chi} becomes unready (lines 20-22), and finally wakes up  the sender thread to notify that the communication is finished.

%When all the above actions are completed, lock \texttt{mutex} is released (line 25).
 
\oomit{
\subsubsection{Output}
The implementation of output $ch!e$ in C is dual to the above case, which is implemented as a function $\texttt{output (tid, cho)}$, where $\texttt{cho}$ is the output channel corresponding to $ch!$.  
\begin{lstlisting}[style=mystyle,language=custom]
void output (int tid, Channel cho) {
  lock mutex;
  channelOutput[cho.channelNo] = tid;
  copyToChannel(ch);
  int i = channelInput[cho.channelNo]; 
  if (i != -1 & (threadState[i] == AVAILABLE || threadState[i] == WAITING_AVAILABLE)) 
  { 
    threadState[i] = cho.channelNo;
    signal cond[i];
    cwait cond[tid] mutex;
    channelOutput[cho.channelNo] = -1;
  }
  else
  {
    threadState[tid] = AVAILABLE;
    localTime[tid] = DBL_MAX;
    updateCurrentTime(tid);
    cwait cond[tid] mutex;
    localTime[tid] = localTime[i];
    threadState[tid] = RUNNING;
    channelOutput[cho.channelNo] = -1;
    signal cond[i];
  }
  unlock mutex;
  }
\end{lstlisting}
At the beginning, the output channel transfers its value to the shared channel content (line 4), and then the two communication cases are presented depending on which among the output and its compatible input occurs first.
Dual to input, we denote  lines 6-15 by \texttt{outAFin(tid, i, cho)}, representing the communication case that output \texttt{cho} of thread \texttt{tid} occurs after its compatible input of thread \texttt{i} and they synchronise with each other, and the contrary case at lines 16-27 by \texttt{outBFin(tid, i, cho)}.

}

\subsubsection{Continuous Evolution}

The C implementation of continuous evolution is \texttt{odec(tid)} as follows. 
It has almost the same control structure as the discretized ODE given in the paper, with $\pwait$ statement implemented using function \texttt{delay}.
%For $wait(h)$, which can be thought of as the smallest unit in the continuous evolution that only cares about time and does not modify all other variables in the corresponding thread, its corresponding translation is \texttt{delay(tid, seconds)}.

\begin{lstlisting}[style=mystyle,language=custom]
void odec(int tid) {
    int i = 1;
    while (i == 1) {
        if (!N(B, eps) || !N^n(B, eps)) {i = 0;}
        if (i == 1) {delay(tid, h); x = x + h*Phi(x,h);} 
        if (currentTime >= T) {i = 0; stop;}
    }
}
\end{lstlisting}

\subsubsection{Continuous Interrupt}

The C implementation is \texttt{odeI (tid, nums, chs)}, followed by an if-then statement on recursive transformation of $p_i$s depending on the return value of \texttt{odeI}. Here \texttt{nums} denotes the number of communication actions in $I$, $\texttt{chs}$ is the array storing these actions.  
It is defined with the help of three auxiliary functions: \texttt{interruptInit}, \texttt{interruptPoll}, and \texttt{interruptClear}.
After each polling operation of duration \texttt{h},  \texttt{x} is updated, followed by two checks: if it reaches the boundary, the communication interrupt terminates directly, indicated by \texttt{ret=-1}; if the execution time exceeds the upper bound \texttt{T}, the whole execution stops. 
\begin{lstlisting}[style=mystyle,language=custom]
void odeI(int tid, int nums, Channel* chs) {
	interruptInit(tid, nums, chs);
	int i = 1, ret = -1;
	while (i == 1) {
		recordTime = localTime[tid];
		ret = interruptPoll(tid, h, nums, chs); 
		if (ret >= 0) {
			i = 0;
			h = localTime[tid] - recordTime;
		} 
		x = x + h*Phi(x,h);
		if (ret < 0) {
			if (!N(B, eps) || !N^n(B, eps)) {
  			i = 0;
				ret = -1;
				interruptClear(tid, nums, chs);
			}
			if (currentTime >= T) {
				i = 0;
				interruptClear(tid, nums, chs);
				stop; 
			}
		}
	}
}
\end{lstlisting}

Function \texttt{interruptInit} initializes actions in \texttt{chs} to be ready, and writes values  to  \texttt{channelContent} buffer from outputs by \texttt{copyToChannel}, and  sets thread state to be \texttt{AVAILABLE}. 

\begin{lstlisting}[style=mystyle,language=custom]
void interruptInit(int tid, int nums, Channel* chs) {
	lock mutex;
	int curChannel;			
	for (int i = 0; i < nums; i++) {
		curChannel = chs[i].channelNo;
		if (chs[i].type == 0) {
			channelInput[curChannel] = tid;
		} else {
			channelOutput[curChannel] = tid;
			copyToChannel(chs[i]);
		}
	}			
	threadState[tid] = AVAILABLE;
	return;
}
\end{lstlisting}

Function \texttt{interruptPoll} checks whether some  communication in \texttt{chs} can occur during $d$ duration. 
\begin{lstlisting}[style=mystyle,language=custom]
int interruptPoll(int tid, double seconds, int nums,
                  Channel* chs) {
  int curChannel, av[nums];
  for (int i = 0; i < nums; i++) {
    curChannel = chs[i].channelNo;
    if (chs[i].type == 0) {
      channelInput[curChannel] = tid;
      m = channelOutput[curChannel];
      if (m != -1 && (threadState[m] == AVAILABLE ||
          threadState[m] == WAITING_AVAILABLE)) { 
        av[k] = i; k++;
      }
      else {
        channelOutput[curChannel] = tid;
        copyToChannel(chs[i]);            
        m = channelInput[curChannel];
        if (m != -1 && (threadState[m] == AVAILABLE ||
            threadState[m] == WAITING_AVAILABLE)) {
          av[k] = i; k++; 
        }
      }
    }
  }
  if (k > 0) {
    u = nondet(0, ..., k-1); 
    i = av[u]; 
    curChannel = chs[i].channelNo;
    if (chs[i].type == 0) { 
      m = channelOutput[curChannel]; 
      inAFout(tid, m, chs[i]);
    }
    else {
      m = channelInput[curChannel]; 
      outAFin(tid, m, chs[i]); 
    }
    clearchannels (chs, nums);
    unlock mutex;
    return av[u];
  }
  localTime[tid] += seconds;
  updateCurrentTime(tid);
  if (localTime[tid] > currentTime) {
    threadState[tid] = WAITING_AVAILABLE;
    cwait cond[tid] mutex;
  }	
  if (threadState[tid] == AVAILABLE) {
    return -1;
  }
  curChannel = threadState[tid];
  int	match_index = -1;
  for (int j = 0; j < nums; j++) {
    if (chs[j].channelNo == curChannel) {
      match_index = j;
    }
  }			
  if (chs[match_index].type == 0) {
    copyFromChannel(chs[match_index]);
    m = channelOutput[curChannel];
  } else {
    m = channelInput[curChannel];
  }
  localTime[tid] = localTime[m];
  threadState[tid] = RUNNING;
  clearchannels(chs, nums);
  signal cond[m];
  unlock mutex;
  return match_index;
}
\end{lstlisting}

Function \texttt{interruptClear} resets the thread state and clear channels. Finally it releases \texttt{mutex} and returns.  
\begin{lstlisting}[style=mystyle,language=custom]
void interruptClear(int tid, int nums, Channel* chs) {
    threadState[tid] = RUNNING;
    clearchannels(chs, nums); 
    unlock mutex;
    return;
}		   
\end{lstlisting}

\end{document}